\documentclass[12pt]{iopart}

\usepackage{graphicx,color,iopams}  
\begin{document}

\title [Understanding MCA based on orbital and quadrupole moments]{Understanding magnetocrystalline anisotropy based on orbital and quadrupole moments}

\author{Yoshio Miura$^{1,2,*}$ \& Jun Okabayashi$^3$}
\address{$^1$Research Center for Magnetic and Spintronic Materials, National Institute for Materials Science (NIMS), Tsukuba 305-0047, Japan}
\address{$^2$Center for Spintronics Research Network (CSRN), Graduate School of Engineering Science, Osaka University, Machikaneyama 1-3, Toyonaka, Osaka 560-8531, Japan}
\address{$^3$Research Center for Spectrochemistry, The University of Tokyo, Bunkyo-ku, Tokyo 113-0033, Japan}
\ead{$^*$MIURA.Yoshio@nims.go.jp}
\vspace{10pt}
\begin{indented}
\item[]June 2022
\end{indented}

\begin{abstract}

Understanding magnetocrystalline anisotropy (MCA) is fundamentally important for developing novel magnetic materials. Therefore, clarifying the relationship between MCA and local physical quantities observed by spectroscopic measurements, such as the orbital and quadrupole moments, is necessary. 
In this review, we discuss MCA and the distortion effects in magnetic materials with transition metals (TMs) based on the orbital and quadrupole moments, which are related to the spin-conserving and spin-flip terms in the second-order perturbation calculations, respectively. 
We revealed that orbital moment stabilized the spin moment  in the direction of the larger orbital moment, while the quadrupole moment stabilized the spin moment along the longitudinal direction of the spin-density distribution. The MCA of the magnetic materials with TMs and their interfaces can be determined from the competition between these two contributions.  We showed that the perpendicular MCA of the face-centered cubic (fcc) Ni with tensile tetragonal distortion arose from the orbital moment anisotropy, whereas that of Mn-Ga alloys originated from the quadrupole moment of spin density.
In contrast, in the Co/Pd(111) multilayer and Fe/MgO(001), both the orbital moment anisotropy and  quadrupole moment of spin density at the interfaces contributed to the perpendicular MCA.
Understanding the MCA of magnetic materials and interfaces based on orbital and quadrupole moments is essential to design MCA of novel magnetic applications.

\end{abstract}

%
%
%
%
%

\section{Introduction }

Magnetocrystalline anisotropy (MCA), in which the internal energy  varies on the magnetization direction, is an important physical property of magnetic materials \cite{2004Sander-JPCM}.
Perpendicular magnetization is required to enhance the thermal stability of magnetization directions in permanent magnets and spintronic materials, used in information storage media and magnetoresistive  random access memory (MRAM) \cite{2017Dieny-RMP, 2017Bhatti-MatToday}.
 Contrary to this, some studies suggest the importance of in-plane MCA of magnetic materials for applications such as a magnetic under-layer of a perpendicular recording medium \cite{2006Hashimoto-JAP} and a spin torque oscillator for microwave-assisted magnetic recording (MAMR) \cite{2010Yoshida-IEEETransMag,2010Igarashi-IEEETransMag}. 
Furthermore, the development of soft magnets requires extremely small MCA under distortion to eliminate losses caused by magnetic hysteresis and circulating loops of current \cite{2022Balakrishna-npjCM}. 
 Hence, understanding MCA is necessary for the development of novel magnetic materials. Several experimental and theoretical studies have been conducted to determine the microscopic origin of MCA.

Theoretically, MCA originates from spin-orbit interaction (SOI), which is a one-body interaction between the electron’s spin and its orbital motion and is caused by the difference in the crystal symmetry because of the change in the magnetization direction via SOI \cite{1996Yosida-SSS}.
MCA of rare earth magnets arises from strong SOI of {\it f} electrons, which has been analyzed by the one-ion Hamiltonian, and understood as a competition between the crystalline electric field and SOI \cite{1991Franse-HMM}. 
On the other hand, the MCA of magnetic materials with TMs should be analyzed using electronic band structures \cite{1991Daalderop-PRB,1994Sakuma-JPSJ,2001Ravindran-PRB,2004Staunton-PRL,2011Sakamaki-APEX,2012JJAP-Kojima,2012Kota-PRB,2013Kotsugi-JMMM,2013Ebert-PRB,2016Ueda-APL}.

First-principles density functional theory (DFT) calculations \cite{1964Hohenberg-PR,1965Kohn-PR,1981Perdew-PRB} are important for quantitatively obtaining MCA energies and electronic structures. However, owing to the small SOI of TMs, understanding the origin of MCA is difficult by simply analyzing the electronic structures by changing the magnetization directions. Therefore, a second-order perturbation analysis of SOI \cite{1989Bruno-PRB,1993Wang-PRB,1994Mathon-PRB,1995Freeman-MSE,1995Wang-JMMM,1996Durr-PRB,1998Laan-JPCM,1999Stohr-JMMM,2006Autes-JPCM} will be useful in deducing the physical origin of MCA.

In the second-order perturbation analysis of MCA, we can resolve MCA energies in terms of atoms, orbitals, and spins.  By expanding the wavefunctions in DFT calculations with local atomic orbitals and calculating the second-order perturbation terms of SOI using the localized basis sets, the contribution of specific atoms, orbitals, and spins  to MCA can be clarified quantitatively \cite{2013Miura-JPCM, 2013Miura-JAP}. In addition,  the spin-conserving and spin-flip terms in the second-order perturbation calculations of SOI are related to the orbital and quadrupole moments observed in spectroscopy, respectively \cite{1989Bruno-PRB,1998Laan-JPCM,1999Stohr-JMMM}. Bruno theoretically showed that the spin-conserving term between occupied and unoccupied minority-spin states is related to the orbital moment  and that MCA energies are proportional to the difference in the orbital moments between the perpendicular and in-plane magnetization directions \cite{1989Bruno-PRB}. This formula, known as the Bruno relation, is widely used to estimate MCA energies from orbital moments obtained by spectroscopic measurements based on X-ray magnetic circular dichroism (XMCD) and X-ray magnetic linear dichroism (XMLD), using the magneto-optical sum rule \cite{1992Thole-PRL,1993Carra-PRL,1998Laan-PRB,1999Laan-PRL}.  
Furthermore, Wang, Wu, and Freeman  showed that the spin-flip term between occupied majority-spin and unoccupied minority-spin states is related to  the quadrupole moment of spin density \cite{1993Wang-PRB}. The quadrupole moment of spin density is the same physical quantity with the magnetic dipole moment,   which is extended to MCA analysis with XMCD and XMLD measurements by van der Laan and St\"{o}hr \cite{1998Laan-JPCM,1999Stohr-JMMM,1995Stohr-PRL}.

According to the Ref. \cite{1998Laan-JPCM}, MCA energies ($E _{\rm MCA}$) in the form of spin-conserving and spin-flip terms are expressed as the orbital moment anisotropy and the magnetic dipole moment:
\begin{equation}
\label{MCA0}
E^{\rm MCA} \approx \frac{\xi}{4} \Delta m_{\rm orb} + \frac{21}{2} \frac{\xi ^2}{\Delta _{\rm exc}} m_{\rm T},
\end{equation}
where $\Delta m_{\rm orb}$ and $m_{\rm T}$ are the orbital moment anisotropy  and magnetic dipole moment, respectively, and $\xi$ and $\Delta _{\rm exc}$ are the SOI constant and exchange splitting of the corresponding materials, respectively.    The orbital moment anisotropy is defined by $\Delta m_{\rm orb}=m_{\rm orb}^{\rm perp.}-m_{\rm orb}^{\rm inp.}$, where $m_{\rm orb}^{\rm perp.}$ and $m_{\rm orb}^{\rm inp.}$ are the orbital moment with the perpendicular and in-plane magnetization, respectively.    

Eq. (\ref{MCA0}) connects MCA energies to orbital and quadrupole moments measured by spectroscopy and is important for the microscopic understanding of MCA in realistic materials.  According to Eq. (\ref{MCA0}), the positive and negative magnetic dipole moments ($m_{\rm T}$) contribute to a perpendicular MCA and an in-plane MCA, respectively. However, there is a lack of intuitive understanding on the relationship between the magnetic dipole moments and MCA, i.e., why the positive (negative) $m_{\rm T}$ contributes to the perpendicular (in-plane) MCA. Because Eq. (1) is based on several approximations and assumptions, separately verifying this equation for various systems  with XMCD and XMLD measurements is necessary. In addition, examining the conformity of theoretical results by the second-order perturbation calculations based on DFT calculations  with experimental results is important. 

In this review, in order to obtain intuitive understandings of MCA, we investigate the dependence of MCA in magnetic materials with TMs on orbital and quadrupole moments based on first-principle DFT and the second-order perturbation calculations,  together with experimental verifications.  First, we show the detailed derivations of the relation between MCA energies and spectroscopic quantities through second-order perturbation calculations of SOI.   The studies of Ref. \cite{1989Bruno-PRB} and \cite{1998Laan-JPCM} have been intended not to give the final answers of MCA, but to give the important direction to understand MCA energy. To give a further important step in the right direction to understand MCA, intuitive understandings and classifications of MCA for various system depending on the interface chemical bonding will be necessary.   
Thereafter, we review collaborative work involving theoretical calculations and XMCD and XMLD measurements for understanding MCA of TM films.
 
We review also strain dependent MCA of bulk and interfaces, because they give rise to the precise determination of MCA characteristics.

First, we discuss the MCA of fcc Ni with in-plane lattice distortion by the piezoelectricity of the underlayer  BaTiO$_3$ \cite{2019Okabayashi-npjQM}. We find that the anisotropy of the orbital moments could describe the MCA of fcc Ni with tetragonal distortion. 
We then discuss the MCA of Mn-Ga alloys, which shows a strong perpendicular MCA without a heavy element. We clarify that the perpendicular MCA of Mn-Ga alloys is attributed to the positive magnetic dipole  of Mn owing to the cigar-type distribution of the spin density \cite{2020Okabayashi-SciRep}. 
Next, we discuss the MCA of the Co($t$)Pd(111) multilayer showing perpendicular MCA when the thickness of the Co layer $t$ is less than 0.6 nm \cite{2018Okabayashi-SciRep}. We reveal that both the interfacial orbital moments anisotropy in Co and magnetic dipole moment in Pd are crucial for the perpendicular MCA. 
Thereafter, we discuss the perpendicular MCA of Fe/MgO(001) \cite{2018Masuda-PRB, 2019Okabayashi-APL}, frequently investigated as an interface MCA of magnetic tunnel junctions (MTJs) used in MRAM applications. We find that tetragonal distortions in MTJs, as well as in interface structures, significantly affect the contribution of the orbital and quadrupole moments to MCA. This dependence of MCA on tetragonal distortions may prove useful for the voltage control of MCA in future spintronic devices \cite{2009Maruyama-NN,2012Kanai-APL,2017Miwa-NC}.  We will give an systematic discussion on MCA of bulk and interfaces and intuitive understandings of MCA based on the orbital and quadrupole moments.

\section{Second-order  perturbation calculations of SOI}

\subsection{MCA energies}

MCA energies are analyzed using second-order perturbation of SOI. The spin-orbit (SO) Hamiltonian is given by 
\[
\hat{H}_{\rm SO}=\sum _I \xi_I    \vec{\hat{L}}\cdot\vec{\hat{S}},
\]
where $\vec{\hat{L}}=(\hat{L}_x,\hat{L}_y,\hat{L}_z)$ and $\vec{\hat{S}}=(\hat{S}_x,\hat{S}_y,\hat{S}_z)$ are the angular momentum and spin angular momentum operators, measured in units of Dirac's constant $\hbar$, respectively; $I$ is the atomic position index, and $\xi _I$ is the spin-orbit coupling strength at $I$. $\xi _I$ has a localized character and is given by 
\begin{equation}
\label{SOI2}
\xi_I \equiv \xi_I(r)=\frac{\hbar ^2}{2m^2c^2}\frac{1}{r}\frac{dV_I(r)}{dr},
\end{equation} 
where $c$ is the speed of light, $m$ is the electron mass, $r$ is the distance from the atomic center, and $V_I(r)$ is the potential between the electron and atomic nucleus \cite{1961Rose-RET}. 
We assume that $H^{(0)}$ is the one-electron Hamiltonian without SOI, satisfying the non-perturbative Schr\"{o}dinger equation
\[
H^{(0)}|\vec{k}n\sigma \rangle=\epsilon_{\vec{k} n \sigma}|\vec{k}n\sigma \rangle,
\]
where $|\vec{k}n\sigma \rangle$ is an unperturbed state of energy $\epsilon_{\vec{k} n \sigma}$ with indices $\vec{k}-$point $\vec{k}$, band $n$, and spin $\sigma$. Using the eigenstate and eigenvalue, the variation in the total energy due to the second-order perturbation of SOI is given by
\begin{equation}
\label{EPT1}
E^{(2)} = -\sum_{\vec{k}} \sum_{n^{\prime}\sigma^{\prime}}^{\rm unocc}  \sum_{n \sigma }^{\rm occ} \frac{|\langle \vec{k} n^{\prime}\sigma^{\prime}|\hat{H}_{\rm SO}|\vec{k}n\sigma\rangle |^2}{\epsilon_{\vec{k} n^{\prime}\sigma^{\prime}}-\epsilon_{\vec{k} n \sigma} }.
\end{equation}
$|\vec{k}n\sigma \rangle$ can be expanded with an orthogonal basis of atomic orbitals labeled $\mu$ (or $\lambda$): 
\begin{equation}
\label{basis}
|\vec{k}n\sigma \rangle=\sum _{j\mu}c_{j\mu \sigma}^{\vec{k}n}|\mu \sigma \rangle e^{i \vec{k}\cdot \vec{R_j}}, 
\end{equation}
where $\vec{R}_j$ is the atomic position at site $j$ in the unit cell. We can obtain the second-order contribution of $\hat{H}_{\rm SO}$ to the total energy as a sum over terms depending on the spin transition processes, atomic orbitals, and atomic sites:
\begin{equation}
\label{EPT2}
E^{(2)}=-\sum_{\sigma \sigma ^{\prime}} \sum_{II^{\prime}} \xi_I \xi_I^{\prime} \sum_{ \lambda \lambda^{\prime} \mu^{\prime} \mu}\langle \lambda \sigma |\vec{\hat{L}}\cdot\vec{\hat{S}}|\lambda^{\prime}\sigma ^{\prime} \rangle \langle \mu^{\prime} \sigma ^{\prime}|\vec{\hat{L}}\cdot\vec{\hat{S}}|\mu \sigma \rangle   G_{II^{\prime}}^{\sigma \sigma ^{\prime}}( \lambda \lambda^{\prime} ; \mu^{\prime} \mu),
\end{equation}
where $G_{II^{\prime}}^{\sigma \sigma ^{\prime}}(\lambda \lambda^{\prime} ; \mu^{\prime} \mu)$ is an integral of joint local density of states (LDOS) given by 
\begin{equation}
\label{JLDOS}
G_{II^{\prime}}^{\sigma \sigma ^{\prime}}(\lambda \lambda^{\prime} ; \mu^{\prime} \mu)=\sum_{\vec{k}}\sum_{n}^{\rm occ} \sum_{n^{\prime}}^{\rm unocc}\frac{c_{I^{\prime}\lambda \sigma}^{\vec{k}n*} c_{I^{\prime}\lambda ^{\prime} \sigma ^{\prime}}^{\vec{k}n^{\prime}}   
 c_{I\mu^{\prime}\sigma^{\prime}}^{\vec{k}n^{\prime}*} c_{I\mu\sigma}^{\vec{k}n} }{\epsilon_{\vec{k} n^{\prime}\sigma^{\prime}}-\epsilon_{\vec{k} n \sigma}}.
\end{equation}
To derive the Eq. (\ref{EPT2}), we assume that SOI acts only at the same atomic site and use the relation $\sum_{jj^{\prime}}e^{i\vec{k}\cdot (\vec{R}_j-\vec{R}_{j^{\prime}})}\xi (|\vec{R}_j-\vec{R}_I|)=\xi _I$. 
 Although SOI is effective within the same atomic site, the joint LDOS $G_{II^{\prime}}^{\sigma \sigma ^{\prime}}(\lambda \lambda^{\prime} ; \mu^{\prime} \mu)$ includes the electronic states for different atomic sites in the unit cell ($I$ and $I^{\prime}$) because of the hybridization of atomic orbitals in the crystal. 
The joint LDOS $G_{II^{\prime}}^{\sigma \sigma ^{\prime}}(\lambda \lambda^{\prime} ; \mu^{\prime} \mu)$ can be calculated using first-principles DFT calculations, and the coefficients $c_{j\mu \sigma}^{\vec{k}n}$ are obtained from the projections of unperturbed eigenstate on each localized atomic orbital. We calculated $G_{II^{\prime}}^{\sigma \sigma ^{\prime}}(\lambda \lambda^{\prime} ; \mu^{\prime} \mu)$ using the Vienna {\it ab initio} simulation package (VASP) code \cite{1993Kresse-PRB,1996Kresse-CMS,1996Kresse-PRB}. Furthermore, we set the spin-orbit coupling constants defined by Eq. (\ref{SOI2}), which are used within VASP code, as follows: $\xi _{{\rm Mn}(3d)}=41.5$ meV , $\xi _{{\rm Fe}(3d)}=54.3$ meV, $\xi _{{\rm Co}(3d)}=69.4$ meV, $\xi _{{\rm Ni}(3d)}=87.2$ meV, $\xi _{{\rm O}(2d)}=24.3$ meV,  $\xi _{{\rm Mg}(2p)}=47.5$ meV, $\xi _{{\rm Ga}(3p)}=35.4$ meV, $\xi _{{\rm Pd}(4d)}=187$ meV.

To obtain MCA energies, its dependence on the magnetization direction must be considered. Noncollinear magnetization can be introduced by expressing the spin-quantum axis of local basis sets $|\sigma \rangle$ ($\sigma = \uparrow$ or $\downarrow$) with a spin-1/2 rotation matrix according to K\"{u}bler's formulation \cite{1988Kubler-JPF,2003Nakamura-PRB}
\begin{eqnarray}
\label{noncollinear}
|\!\uparrow \rangle = e^{i\frac{\phi}{2}}\cos \frac{\theta}{2} |\tilde{\uparrow} \rangle +e^{-i\frac{\phi}{2}}\sin \frac{\theta}{2} |\tilde{\downarrow} \rangle, \\
\label{noncollinear2}
|\!\downarrow \rangle = -e^{i\frac{\phi}{2}}\sin \frac{\theta}{2} |\tilde{\uparrow} \rangle+e^{-i\frac{\phi}{2}}\cos \frac{\theta}{2} |\tilde{\downarrow} \rangle,
\end{eqnarray}
where $\theta$ and $\phi$ are the polar and azimuthal angles of magnetization with respect to the $z$ and $x$-axes of the system, and $z$ axis is perpendicular to the plane of the tetragonal or hexagonal systems. $|\tilde{\sigma} \rangle$ indicates a spin state along the global spin-quantum axis, fixed to the $z$ axis of the system. 

By using Eqs. (\ref{noncollinear}) and (\ref{noncollinear2}) in the matrix  element $\langle \mu^{\prime} \sigma ^{\prime}|\vec{\hat{L}} \cdot \vec{\hat{S}}|\mu \sigma \rangle$, we obtain the following expressions depending on the magnetization direction:
\begin{eqnarray}
\langle \mu^{\prime}\! \uparrow\! |\vec{\hat{L}}\cdot\vec{\hat{S}}|\mu\! \uparrow \rangle=\frac{1}{2}[\sin\theta\cos\phi\langle \mu^{\prime} |\hat{L}_x|\mu \rangle \nonumber \\ 
~~~~~~~~~~~~~~~~~~~~~~~~-\sin\theta\sin\phi\langle \mu^{\prime} |\hat{L}_y|\mu \rangle+\cos\theta\langle \mu^{\prime} |\hat{L}_z|\mu \rangle], \nonumber \\
\langle \mu^{\prime}\! \downarrow\! |\vec{\hat{L}}\cdot\vec{\hat{S}}|\mu \! \downarrow \rangle=-\frac{1}{2}[\sin\theta\cos\phi\langle \mu^{\prime} |\hat{L}_x|\mu \rangle \nonumber \\
~~~~~~~~~~~~~~~~~~~~~~~~~~-\sin\theta\sin\phi\langle \mu^{\prime} |\hat{L}_y|\mu \rangle+\cos\theta\langle \mu^{\prime} |\hat{L}_z|\mu \rangle], \nonumber \\
\langle \mu^{\prime}\! \uparrow\! |\vec{\hat{L}}\cdot\vec{\hat{S}}|\mu \! \downarrow \rangle=\frac{1}{2}[(\cos\theta\cos\phi -i\sin\phi)\langle \mu^{\prime} |\hat{L}_x|\mu \rangle \nonumber \\
~~~~~~~~~~~~~~~~~~~~~~~~-(\cos\theta\sin\phi +i\cos\phi)\langle \mu^{\prime} |\hat{L}_y|\mu \rangle -\sin\theta\langle \mu^{\prime} |\hat{L}_z|\mu \rangle], \nonumber \\
\langle \mu^{\prime}\! \downarrow\! |\vec{\hat{L}}\cdot\vec{\hat{S}}|\mu \! \uparrow \rangle=\frac{1}{2}[(\cos\theta\cos\phi +i\sin\phi)\langle \mu^{\prime} |\hat{L}_x|\mu \rangle \nonumber \\
~~~~~~~~~~~~~~~~~~~~~~~~-(\cos\theta\sin\phi -i\cos\phi)\langle \mu^{\prime} |\hat{L}_y|\mu \rangle -\sin\theta\langle \mu^{\prime} |\hat{L}_z|\mu \rangle], \nonumber
\end{eqnarray}
where we have used the eigenstate and eigenvalue relation between $\vec{\hat{S}}=(\hat{S}_x,\hat{S}_y,\hat{S}_z)$ and $|\tilde{\sigma} \rangle$ and the orthonormal property of the eigenstate $\langle \tilde{\sigma} |\tilde{\sigma} ^{\prime} \rangle =\delta _{\tilde{\sigma} \tilde{\sigma} ^{\prime}}$. Because we consider the uniaxial MCA energies, the energy difference between the perpendicular magnetization ($\theta=0$ and $\phi=0$) and in-plane magnetization ($\theta=\frac{\pi}{2}$ and $\phi=0$) should be calculated. In this case, the matrix elements of SOI for each magnetization direction are given by
\begin{eqnarray}
\label{spinol1}
\langle \mu^{\prime}\! \uparrow\! |\vec{\hat{L}}\cdot\vec{\hat{S}}|\mu \! \uparrow \rangle _{\theta=0,\phi=0} =-\langle \mu^{\prime}\! \downarrow\! |\vec{\hat{L}}\cdot\vec{\hat{S}}|\mu \! \downarrow \rangle _{\theta=0,\phi=0} = \frac{1}{2} \langle \mu^{\prime}  |\hat{L}_z |\mu \rangle, \\
\label{spinol2}
\langle \mu^{\prime}\! \uparrow\! |\vec{\hat{L}}\cdot\vec{\hat{S}}|\mu \! \uparrow \rangle _{\theta=\frac{\pi}{2},\phi=0} =-\langle \mu^{\prime} \! \downarrow\! |\vec{\hat{L}}\cdot\vec{\hat{S}}|\mu \! \downarrow \rangle _{\theta=\frac{\pi}{2},\phi=0} = \frac{1}{2} \langle \mu^{\prime}  |\hat{L}_x |\mu \rangle, \\
\label{spinol3}
\langle \mu^{\prime}\! \uparrow\! |\vec{\hat{L}}\cdot\vec{\hat{S}}|\mu \! \downarrow \rangle _{\theta=0,\phi=0} =\langle \mu^{\prime}\! \downarrow\! |\vec{\hat{L}}\cdot\vec{\hat{S}}|\mu \! \uparrow \rangle ^* _{\theta=0,\phi=0} = \frac{1}{2} \langle \mu^{\prime}  |\hat{L}_x-i\hat{L}_y |\mu \rangle, \\
\label{spinol4}
\langle \mu^{\prime}\! \uparrow\! |\vec{\hat{L}}\cdot\vec{\hat{S}}|\mu \! \downarrow \rangle _{\theta=\frac{\pi}{2},\phi=0} =\langle \mu^{\prime}\! \downarrow\! |\vec{\hat{L}}\cdot\vec{\hat{S}}|\mu \!\uparrow  \rangle ^* _{\theta=\frac{\pi}{2},\phi=0} = -\frac{1}{2} \langle \mu^{\prime}  |\hat{L}_z+i\hat{L}_y |\mu \rangle.
\end{eqnarray}
Since we define the MCA energy as positive for perpendicular magnetization ($\theta=0$ and $\phi=0$), the MCA energy in the second-order perturbation calculations is given by
\begin{eqnarray}
~~~~~~~~~~~~~~E_{\rm MCA}^{(2)}\equiv E^{(2)}(\theta=\frac{\pi}{2},\phi=0)-E^{(2)}(\theta=0,\phi=0) \nonumber \\
\label{MCA2}
~~~~~~~~~~~~~~~~~~~~~~=E_{\rm MCA}^{\uparrow\uparrow}+E_{\rm MCA}^{\downarrow\downarrow}+E_{\rm MCA}^{\uparrow\downarrow}+E_{\rm MCA}^{\downarrow\uparrow} \\
\label{MCA3}
=\sum_{II^{\prime}}\frac{\xi _I \xi_{I^{\prime}}}{4} 
\!\!\sum_{\lambda \lambda^{\prime} \mu \mu^{\prime}} [ \langle \lambda  |\hat{L}_z | \lambda ^{\prime} \rangle \langle \mu^{\prime}  |\hat{L}_z |\mu \rangle - \langle \lambda  |\hat{L}_x |\lambda ^{\prime} \rangle \langle \mu^{\prime}  |\hat{L}_x | \mu \rangle ] G_{II^{\prime}}^{\uparrow \uparrow}(\lambda \lambda^{\prime} ; \mu^{\prime} \mu)  \nonumber \\
~~~~~~~~~~+\!\!\sum_{\lambda \lambda^{\prime} \mu \mu^{\prime}} [ \langle \lambda  |\hat{L}_z | \lambda ^{\prime} \rangle \langle \mu^{\prime}  |\hat{L}_z |\mu \rangle - \langle \lambda  |\hat{L}_x |\lambda ^{\prime} \rangle \langle \mu^{\prime}  |\hat{L}_x | \mu \rangle ] G_{II^{\prime}}^{\downarrow \downarrow}(\lambda \lambda^{\prime} ; \mu^{\prime} \mu)  \nonumber \\
~~~~~~~~~~-\!\!\sum_{\lambda \lambda^{\prime} \mu \mu^{\prime}} [ \langle \lambda  |\hat{L}_z | \lambda ^{\prime} \rangle \langle \mu^{\prime}  |\hat{L}_z |\mu \rangle - \langle \lambda  |\hat{L}_x |\lambda ^{\prime} \rangle \langle \mu^{\prime}  |\hat{L}_x | \mu \rangle ] G_{II^{\prime}}^{\uparrow \downarrow}(\lambda \lambda^{\prime} ; \mu^{\prime} \mu)  \nonumber \\
~~~~~~~~~~-\!\!\sum_{\lambda \lambda^{\prime} \mu \mu^{\prime}} [ \langle \lambda  |\hat{L}_z | \lambda ^{\prime} \rangle \langle \mu^{\prime}  |\hat{L}_z |\mu \rangle - \langle \lambda  |\hat{L}_x |\lambda ^{\prime} \rangle \langle \mu^{\prime}  |\hat{L}_x | \mu \rangle ] G_{II^{\prime}}^{\downarrow \uparrow}(\lambda \lambda^{\prime} ; \mu^{\prime} \mu).
\end{eqnarray}
In $E_{\rm MCA}^{(2)}$, $\langle \hat{L}_y \rangle$ terms in Eqs. (\ref{spinol3}) and (\ref{spinol4}) are automatically canceled out. $E_{\rm MCA}^{\sigma\sigma ^{\prime}}$ is the spin-resolved MCA energy, where $E_{\rm MCA}^{\uparrow \uparrow}$ and $E_{\rm MCA}^{\downarrow \downarrow}$ are spin-conserving terms, and $E_{\rm MCA}^{\uparrow \downarrow}$ and $E_{\rm MCA}^{\downarrow \uparrow}$ are spin-flip terms. It is important to note that the sign of the matrix elements of $\hat{L}_z$ and $\hat{L}_x$ are different for the spin-conserving (positive) and spin-flip (negative) terms. This means that the local electronic structures around the Fermi level have an opposite contribution to MCA in the spin-conserving and spin-flip processes. The opposite sign between the spin-conserving and the spin-flip terms arises from the eigenvalue of $\hat{S}_z$ for the spin eigenstate $|\sigma\rangle$, i.e., $\hat{S}_z |\uparrow \rangle = +\frac{1}{2}  |\uparrow \rangle$ and $\hat{S}_z |\downarrow \rangle = -\frac{1}{2}  |\uparrow \rangle$.  The spin-conserving term includes two eigenvalues with the same sign, wheres the spin-flip term has two eigenvalues with the different sign, leading the opposite contribution to MCA.
\renewcommand{\arraystretch}{1.5}
\begin{table}[tp]
\centering
\caption{\label{Matrix}Nonzero matrix elements of the angular momentum operator \\ $\vec{\hat{L}}=(\hat{L}_x,\hat{L}_y,\hat{L}_z)$ for the $d$ orbitals in Cartesian coordinates.}
\footnotesize
\begin{tabular}{cc|cc|c}
\br
$\langle d_{x^2-y^2}  |\hat{L}_x | d_{yz} \rangle=1$ && $\langle d_{xy}  |\hat{L}_x | d_{zx} \rangle=1$ && $\langle d_{3z^2-r^2}  |\hat{L}_x | d_{yz} \rangle =\sqrt{3}$ \\
$\langle d_{x^2-y^2}  |\hat{L}_y | d_{zx} \rangle=1$ && $\langle d_{yz}  |\hat{L}_y | d_{xy} \rangle=1$ && $\langle d_{3z^2-r^2}  |\hat{L}_y | d_{zx} \rangle =\sqrt{3}$ \\
$\langle d_{x^2-y^2}  |\hat{L}_z | d_{xy} \rangle=2$ && $\langle d_{zx}  |\hat{L}_z | d_{yz} \rangle=1$ &&  \\
\br

\end{tabular}
\end{table}
\normalsize

 In Table \ref{Matrix}, we show the nonzero matrix elements of the angular momentum operator $\vec{\hat{L}}=(\hat{L}_x,\hat{L}_y,\hat{L}_z)$ for the $d$ orbitals in Cartesian coordinates. The matrix elements of $\hat{L}_z$ are nonzero for $d$ orbitals with the same magnetic quantum numbers for occupied and unoccupied states, such as $d_{xy}-d_{x^2-y^2}$ and $d_{yz}-d_{zx}$, contributing to the perpendicular MCA in the spin-conserving terms ($E_{\rm MCA}^{\uparrow\uparrow}$ and $E_{\rm MCA}^{\downarrow\downarrow}$) and in-plane MCA in the spin-flip terms ($E_{\rm MCA}^{\uparrow\downarrow}$ and $E_{\rm MCA}^{\downarrow\uparrow}$). On the other hand, the matrix elements of $\hat{L}_x$ are nonzero for $d$ orbitals when the difference in magnetic quantum numbers between the occupied and unoccupied states is $\pm 1$, contributing to the in-plane MCA in $E_{\rm MCA}^{\uparrow\uparrow}$ and $E_{\rm MCA}^{\downarrow\downarrow}$ terms and the perpendicular MCA in $E_{\rm MCA}^{\uparrow\downarrow}$ and $E_{\rm MCA}^{\downarrow\uparrow}$ terms. These analyses enable us understand the relationship between the local electronic structures around the Fermi level and their MCA contribution at each atomic site; This may be useful in designing new ferromagnetic materials with a strong perpendicular MCA for spintronics applications.

\subsection{Orbital moments}

The first-order correction of wavefunctions for perturbation of SOI is given by
\begin{equation}
\label{first_wave}
|\vec{k}n\sigma\rangle_{(1)} = |\vec{k}n\sigma\rangle +\sum_{n^{\prime}\sigma^{\prime}}^{\rm unocc}  \sum_{n \sigma }^{\rm occ} \frac{\langle \vec{k} n^{\prime}\sigma^{\prime}|\hat{H}^{\rm SO}|\vec{k}n\sigma\rangle }{\epsilon_{\vec{k} n^{\prime}\sigma^{\prime}}-\epsilon_{\vec{k} n \sigma} }|\vec{k}n^{\prime}\sigma^{\prime}\rangle.
\end{equation}
The orbital moment $ \langle \hat{L}_{\zeta} \rangle $, which is the component of $\vec{\hat{L}}$ parallel to the direction of the spin quantum axis $\zeta =(\theta,\phi$), is given by as the expectation value of the angular momentum operator in Eq. (\ref{first_wave}). Because the expectation value of $\hat{L}_{\zeta}$ for the unperturbed states is zero, the orbital moment is given by
\[
\langle \hat{L}_{\zeta} \rangle\simeq2\sum_{\vec{k}}\sum_{n^{\prime}\sigma^{\prime}}^{\rm unocc}  \sum_{n \sigma }^{\rm occ} \langle \vec{k} n\sigma |\hat{L}_{\zeta} |\vec{k}n^{\prime}\sigma^{\prime}\rangle \frac{\langle \vec{k} n^{\prime}\sigma^{\prime}|\hat{H}^{\rm SO}|\vec{k}n\sigma\rangle }{\epsilon_{\vec{k} n^{\prime}\sigma^{\prime}}-\epsilon_{\vec{k} n \sigma} },
\]
where the factor of 2 comes from the Hermitian conjugate of the SOI Hamiltonian. We neglect the squared term of $\hat{H}^{\rm SO}$.

By expanding $|\vec{k}n\sigma \rangle$ with the orthogonal basis given by Eq. (\ref{basis}), we obtained the orbital moment at atomic site $I$ using the following equation:
\[
\langle \hat{L}_{\zeta} \rangle _I=-2\xi _I \sum_{\sigma \sigma^{\prime}} \sum_{I^{\prime}}\!\!\sum_{\lambda \lambda^{\prime}\mu^{\prime}\mu}   \langle \lambda \sigma |\hat{L}_{\zeta} |\lambda^{\prime}\sigma^{\prime}\rangle \langle  \mu^{\prime}\sigma^{\prime}|\vec{\hat{L}}\cdot\vec{\hat{S}}|\mu \sigma\rangle G_{II^{\prime}}^{\sigma \sigma ^{\prime}}(\lambda \lambda^{\prime} ; \mu^{\prime} \mu).
\]
We note that $\langle \hat{L}_{\zeta} \rangle =\sum_I \langle \hat{L}_{\zeta} \rangle _I$. The sign of the orbital moment is determined by the term $\langle \lambda \sigma |\hat{L}_{\zeta} |\lambda^{\prime}\sigma^{\prime}\rangle$, which  is equivalent to the $\vec{\hat{L}}$ component parallel to the direction of the spin quantum axis, and the following relation holds \cite{1989Bruno-PRB}:
\begin{equation}
\label{orb_soi_relation}
\langle \lambda \sigma |\hat{L}_{\zeta} |\lambda^{\prime}\sigma^{\prime}\rangle=2{\rm sgn}(\sigma)\delta _{\sigma \sigma^{\prime}}\langle \lambda \sigma |\vec{\hat{L}}\cdot\vec{\hat{S}} |\lambda^{\prime}\sigma^{\prime}\rangle,
\end{equation}
where ${\rm sgn}(\uparrow)=+1$ and ${\rm sgn}(\uparrow)=-1$. Due to $\delta _{\sigma \sigma^{\prime}}$ in Eq. (\ref{orb_soi_relation}), the spin-flip terms do not contribute to orbital moment. Thus, the orbital moment can be written only by the spin-conserving terms as follow:
\[
\langle \hat{L}_{\zeta} \rangle _I=-4\xi _I \!\! \sum_{\sigma }{\rm sgn}(\sigma)\!\! \sum_{I^{\prime}}\!\! \sum_{\lambda \lambda^{\prime}\mu^{\prime}\mu} \!\! \langle \lambda \sigma |\vec{\hat{L}}\cdot\vec{\hat{S}} |\lambda^{\prime}\sigma\rangle \langle  \mu^{\prime}\sigma|\vec{\hat{L}}\cdot\vec{\hat{S}}|\mu \sigma\rangle G_{II^{\prime}}^{\sigma \sigma }(\lambda \lambda^{\prime} ; \mu^{\prime} \mu).
\]
Finally, the orbital moment with the spin moment along the $z$ axis ($\theta=0$ and $\phi=0$) corresponding to $m_{\rm orb}^{\rm perp.}$ and the spin moment along the $x$ axis ($\theta=\frac{\pi}{2}$ and $\phi=0$) corresponding to $m_{\rm orb}^{\rm inp.}$ are given by 
\begin{eqnarray}
\langle \hat{L}_z \rangle _I=\xi _I \sum_{I^{\prime}} \sum_{\lambda \lambda^{\prime}\mu^{\prime}\mu}  \langle \lambda |\hat{L}_z |\lambda^{\prime}\rangle \langle  \mu^{\prime}|\hat{L} _z |\mu \rangle [G_{II^{\prime}}^{\downarrow \downarrow }(\lambda \lambda^{\prime} ; \mu^{\prime} \mu)-G_{II^{\prime}}^{\uparrow \uparrow }(\lambda \lambda^{\prime} ; \mu^{\prime} \mu)], \nonumber  \\
\langle \hat{L}_x \rangle _I=\xi _I \sum_{I^{\prime}} \sum_{\lambda \lambda^{\prime}\mu^{\prime}\mu}  \langle \lambda |\hat{L}_x |\lambda^{\prime}\rangle \langle  \mu^{\prime}|\hat{L} _x |\mu \rangle [G_{II^{\prime}}^{\downarrow \downarrow }(\lambda \lambda^{\prime} ; \mu^{\prime} \mu)-G_{II^{\prime}}^{\uparrow \uparrow }(\lambda \lambda^{\prime} ; \mu^{\prime} \mu)]. \nonumber
\end{eqnarray}
To obtain these equations, we used the relation in Eqs. (\ref{spinol1})–(\ref{spinol2}), and the eigenstates and eigenvalues of the spin angular momentum operator.

If we assume that the spin-flip terms $E_{\rm MCA}^{\uparrow\downarrow}$ and $E_{\rm MCA}^{\downarrow\uparrow}$ and the spin-conserving term $E_{\rm MCA}^{\uparrow\uparrow}$ in Eq. (\ref{MCA3}) are negligible compared to the spin-conserving term $E_{\rm MCA}^{\downarrow\downarrow}$, the MCA energy can be expressed as the sum of the orbital moment anisotropy at each atomic site $\Delta m_{\rm orb}^I$:
\begin{equation}
\label{morbdiff}
E_{\rm MCA}^{(2)}\approx \frac{1}{4}\sum_I \xi_I[\langle \hat{L}_z \rangle _I-\langle \hat{L}_x \rangle _I]=\frac{1}{4}\sum_I \xi_I \Delta m_{\rm orb}^I, 
\end{equation}
This is the Bruno relation, commonly used to connect MCA energies and the observed orbital moments by XMCD.

Because the ferromagnetic  materials with more than half elements as Fe, Co, and Ni have fully occupied majority-spin states, the unoccupied majority-spin states around the Fermi level are negligible. In this case, neglecting $E_{\rm MCA}^{\uparrow\uparrow}$ and $E_{\rm MCA}^{\downarrow\uparrow}$ is reasonable for MCA energies. However, there are cases in which the spin-flip-term through the unoccupied minority-spin states $E_{\rm MCA}^{\uparrow\downarrow}$ is not too small to be ignored compared to the spin-conserving term $E_{\rm MCA}^{\downarrow\downarrow}$, and the Bruno relation fails to describe MCA energies, both quantitatively and qualitatively. If the energy depth of the occupied majority-spin states is far from the Fermi level owing to strong exchange splitting, we can neglect the spin-flip term $E_{\rm MCA}^{\uparrow\downarrow}$ because of the large denominator in Eq. (\ref{JLDOS}), and the Bruno relation well describes MCA energies.

\subsection{Quadrupole moments}

Wang, Wu, and Freeman proposed that the spin-flip term $E_{\rm MCA}^{\uparrow\downarrow}$ can be related to quadrupole moments \cite{1993Wang-PRB}. We start the formulation of MCA energies in Eq. (\ref{EPT1}) in the second-order perturbation. We then expand the unperturbed unoccupied eigenstates $|\vec{k}n^{\prime} \downarrow \rangle$ using the local atomic orbitals of Eq. (\ref{basis}) and retain the occupied eigenstates $|\vec{k}n \uparrow \rangle$.
\begin{eqnarray}
\label{MCA4}
E^{\rm \uparrow \downarrow}_{\rm MCA} = -\frac{1}{4} \sum_{\vec{k}} \sum_{n^{\prime}}^{\rm unocc}  \sum_{n }^{\rm occ} \sum_I \xi_I^2 \Biggr[\frac{\sum _{\lambda^{\prime} \mu^{\prime}} c_{I\lambda^{\prime} \downarrow}^{\vec{k}n^{\prime}*}c_{I\mu^{\prime} \downarrow}^{\vec{k}n^{\prime}} \langle \lambda^{\prime}\!\! \downarrow |\hat{L}_z|\vec{k}n \!\! \uparrow \rangle \langle \vec{k} n \!\! \uparrow|\hat{L}_z| \mu^{\prime} \!\! \downarrow \rangle }{\epsilon_{\vec{k} n^{\prime} \downarrow}-\epsilon_{\vec{k} n \uparrow}}  \nonumber \\
~~~~~~~~~~~~~~~~~~~~~~~~~~~-\frac{\sum _{\lambda^{\prime} \mu^{\prime}} c_{I\lambda^{\prime} \downarrow}^{\vec{k}n^{\prime}*}c_{I\mu^{\prime} \downarrow}^{\vec{k}n^{\prime}} \langle \lambda^{\prime} \!\! \downarrow |\hat{L}_x|\vec{k}n \!\! \uparrow \rangle \langle \vec{k} n \!\! \uparrow|\hat{L}_x|\mu^{\prime} \!\! \downarrow \rangle }{\epsilon_{\vec{k} n^{\prime} \downarrow}-\epsilon_{\vec{k} n \uparrow}}\Biggr].
\end{eqnarray}
The factor of $\frac{1}{4}$ comes from the eigenvalues of two spin angular momentum operators, and we leaves brakets of spin states $|\sigma \rangle$ to clearly specify the spin state in the second-order perturbation term. Here, we express a spin state by $\sigma$ for the global spin-quantum axis along $z$-axis. 
Wang, Wu, and Freeman introduced two approximations to relate the spin-flip term and quadrupole moment.
 First, the replacement of the difference in eigenvalues between the unoccupied and occupied states in the denominator with $\Delta _{\rm exc}$ is considered:
\begin{equation} 
\label{exchange_splitting}
\epsilon_{\vec{k} n^{\prime} \downarrow}-\epsilon_{\vec{k} n \uparrow} \approx \Delta _{\rm exc},
\end{equation} 
where $\Delta _{\rm exc}$ is the exchange splitting of ferromagnetic materials, corresponding to the range of eigenvalues between the majority-spin occupied and minority-spin unoccupied states. Second, the use of the completeness relation of brakets on occupied majority-spin states:
\begin{equation}
\label{completeness}
\sum _n ^{\rm occ} | n \uparrow \rangle \langle n \uparrow | \approx 1.
\end{equation}
This relation holds if we consider ferromagnetic materials with more than half elements as Fe, Co, and Ni, where the majority-spin states are fully occupied. In this case, the sum of all occupied states is equivalent to the sum of all states.

By using Eqs. (\ref{exchange_splitting}) and (\ref{completeness}), the square of the expectation of the angular momentum operator can be expressed as the expectation of the square of the angular momentum operator:
\[
E^{\rm \uparrow \downarrow}_{\rm MCA} \approx -\frac{1}{4} \sum_{\vec{k}} \sum_{n^{\prime}}^{\rm unocc} \sum_I \xi_I^2  \frac{\sum _{\lambda^{\prime}  \mu^{\prime}} c_{I\lambda^{\prime} \downarrow}^{\vec{k}n^{\prime}*}c_{I\mu^{\prime} \downarrow}^{\vec{k}n^{\prime}} [ \langle \lambda^{\prime} \!\! \downarrow |\hat{L}_z^2|\mu^{\prime} \!\! \downarrow \rangle -\langle \lambda^{\prime} \!\! \downarrow |\hat{L}_x^2|\mu^{\prime} \!\! \downarrow \rangle ]}{\Delta _{\rm exc}} .
\]
Furthermore, if we assume the tetragonal system which has the same lattice structure for the in-plane $x$ and $y$ axes, the following relation holds.
\[
\langle  \hat{L}_x^2 \rangle=\langle  \hat{L}_y^2 \rangle=\frac{1}{2}\langle  \hat{L}_x^2+\hat{L}_y^2 \rangle=\frac{1}{2} \langle  \hat{L}^2-\hat{L}_z^2\rangle.
\]

Thus, for tetragonal systems, we have
\[
E^{\rm \uparrow \downarrow}_{\rm MCA} \approx -\frac{1}{8} \sum_I \xi_I^2 \sum_{\vec{k}} \sum_{n^{\prime}}^{\rm unocc}  \frac{\sum _{\mu^{\prime}} |c_{I\mu^{\prime} \downarrow}^{\vec{k}n^{\prime}}|^2 \langle \mu^{\prime} \downarrow |3\hat{L}_z^2-\hat{L}^2|\mu^{\prime} \downarrow \rangle }{\Delta _{\rm exc}}. \nonumber
\]
Here, the matrix of $\langle  \hat{L}^2 \rangle$ and $\langle \hat{L}_z^2\rangle$ are diagonal for atomic-orbital basis set, and we use the orthonormal property, $\langle \lambda ^{\prime} |\mu ^{\prime} \rangle =\delta _{\lambda ^{\prime} \mu ^{\prime}}$. The operator $3\hat{L}_z^2-\hat{L}^2$ has the same form as the z-component of  intra-atomic  quadrupole moment operator:
\begin{equation}
\label{qzz}
\hat{Q}_{zz}\equiv \frac{2}{21}(3\hat{L}_z^2-\hat{L}^2).
\end{equation}
Finally, we can express the spin-flip term of MCA energies related to the quadrupole moments of the unoccupied  minority-spin electron densities
\begin{equation}
\label{MCA7}
E^{\rm \uparrow \downarrow}_{\rm MCA} \approx -\frac{21}{16} \sum_I \xi_I^2 \sum_{\vec{k}} \sum_{n^{\prime}}^{\rm unocc}  \frac{ \sum _{\mu^{\prime}} |c_{I\mu^{\prime} \downarrow}^{\vec{k}n^{\prime}}|^2 \langle \mu^{\prime} \downarrow |\hat{Q}_{zz}|\mu^{\prime} \downarrow \rangle }{\Delta _{\rm exc}}.
\end{equation}
This is an approximate expression of the spin-flip term $E^{\rm \uparrow \downarrow}_{\rm MCA}$ related to the quadrupole moment. 
Based on the definition of quadrupole moment in Eq. (\ref{qzz}), oblate distributions of unoccupied minority-spin electrons along $z$ axis (cigar-like quadrupole), e.g., $d_{3z^2-r^2}$ orbital, gives a negative quadrupole moment, yielding perpendicular (positive) MCA through Eq. (\ref{MCA7}). In contrast, prolate distributions of unoccupied minority-spin electrons  along $xy$ plane (pancake-like quadrupole), e.g., $d_{x^2-y^2}$ and $d_{xy}$ orbitals, produce a positive quadrupole moment, yielding an in-plane (negative) MCA.

In Eq. (\ref{MCA7}), the spin-flip term of MCA energies $E^{\rm \uparrow \downarrow}_{\rm MCA}$ is described by the quadrupole moments of the unoccupied minority-spin electrons. In addition, $E^{\rm \uparrow \downarrow}_{\rm MCA}$ can be expressed by the quadrupole moments of occupied minority-spin electrons :
\begin{equation}
\label{MCA8}
E^{\rm \uparrow \downarrow}_{\rm MCA} \approx +\frac{21}{16} \sum_I \xi_I^2 \sum_{\vec{k}} \sum_{n}^{\rm occ}  \frac{\sum _{\mu} |c_{I\mu \downarrow}^{\vec{k}n}|^2 \langle \mu \downarrow |\hat{Q}_{zz}|\mu \downarrow \rangle }{\Delta _{\rm exc}},
\end{equation}
where $\sum _{n^{\prime}}^{\rm unocc}=\sum _{n}^{\rm all}-\sum _{n}^{\rm occ}$ and the quadrupole moment $\langle \hat{Q}_{zz} \rangle_I^{\downarrow}$ including all states (occupied  and unoccupied states) is zero, and $\langle \hat{Q}_{zz} \rangle_I^{\downarrow}=\sum_{\vec{k}} \sum_{n} \sum _{\mu} |c_{I\mu \downarrow}^{\vec{k}n}|^2 \langle  \mu \!\! \downarrow \!\!|\hat{Q}_{zz}|\mu \!\! \downarrow \rangle$.

Eq. (\ref{MCA7}) and Eq. (\ref{MCA8}) shows that the contribution of quadrupole moments to MCA energies is opposite between the occupied and unoccupied states in the minority-spin electrons. Then, we considered the quadrupole moments of spin density  corresponding to the magnetic dipole moment $m_{\rm T}$. Again, we assumed that magnetic materials with more than half elements as Fe, Co, and Ni have fully occupied majority-spin states. This provides the following relationship:  
\[
\sum_{\vec{k}} \sum _n^{\rm occ} \sum _{\mu} |c_{I\mu \uparrow}^{\vec{k}n}|^2 \langle \mu \uparrow |\hat{Q}_{zz}|\mu \uparrow \rangle \approx \sum_{\vec{k}}  \sum _n^{\rm all} \sum _{\mu} |c_{I\mu \uparrow}^{\vec{k}n}|^2 \langle \mu \uparrow |\hat{Q}_{zz}|\mu \uparrow \rangle =0.
\]
Thus, we can write Eq. ~ (\ref{MCA8}) using a  intra-atomic  magnetic dipole moment:
\begin{eqnarray}
\label{MCA9}
E^{\rm \uparrow \downarrow}_{\rm MCA} \approx -\frac{21}{16} \! \sum_I \! \frac{\xi_I^2}{\Delta _{\rm exc}} \! \sum_{\vec{k}} \sum_{n}^{\rm occ} \sum _{\mu}  [|c_{I\mu \uparrow}^{\vec{k}n}|^2\langle \mu \!\! \uparrow \!|\hat{Q}_{zz}|\mu \!\! \uparrow \rangle- |c_{I\mu \downarrow}^{\vec{k}n}|^2\langle \mu \!\! \downarrow \!\! |\hat{Q}_{zz}|\mu \! \downarrow \rangle ] \nonumber \\
~~~~~~~~=-\frac{21}{8} \! \sum_I \! \frac{\xi_I^2}{\Delta _{\rm exc}} \! \sum_{\vec{k}} \sum_{n}^{\rm occ} \sum _{\mu}  \sum _{\sigma} |c_{I\mu \sigma}^{\vec{k}n}|^2\langle \mu \sigma |\hat{Q}_{zz}\hat{S}_z|\mu  \sigma \rangle  \nonumber \\
~~~~~~~~=+\frac{21}{8} \! \sum_I \! \frac{\xi_I^2}{\Delta _{\rm exc}} \! m_{\rm T}^I. 
\end{eqnarray}
where the magnetic dipole moment and quadrupole moment of the spin density at each atomic site is related by following equation:
\begin{equation}
\label{mtq}
m_{\rm T}^I \equiv - \langle \hat{Q}_{zz}\hat{S}_z \rangle _I =-\frac{1}{2}[\langle \hat{Q}_{zz} \rangle^{\uparrow}_I -\langle \hat{Q}_{zz} \rangle^{\downarrow}_I].
\end{equation}
Furthermore, we can rewrite the intra-atomic magnetic dipole moment using the spin moment projected to each atomic orbitals $m_{\rm spin}^{\mu}$ as follow:
\begin{eqnarray}
m_{\rm T}^{I}=-\frac{1}{21}(m_{\rm spin}^{p_x}+m_{\rm spin}^{p_y}-2m_{\rm spin}^{p_z} \nonumber \\
\label{mtspin}
~~~~~~~~~~~~~~~~+6m_{\rm spin}^{d_{x^2-y^2}}+6m_{\rm spin}^{d_{xy}}-3m_{\rm spin}^{d_{yz}}-3m_{\rm spin}^{d_{zx}}-6m_{\rm spin}^{d_{3z^2-r^2}}).
\end{eqnarray}
The  intra-atomic  magnetic dipole moment $m^I_{\rm T}$ can be observed by XMCD and XMLD measurements \cite{1995Stohr-PRL}, and a positive (negative) $m_{\rm T}^I$ indicates the contribution of the spin-flip term to the perpendicular (in-plane) MCA.
Because the $m^I_{\rm T}$ term is directly related to MCA energies, we do not need to consider the anisotropy of $m^I_{\rm T}$ to discuss its contribution to MCA, unlike the orbital moments in Eq.  (\ref{morbdiff}).  Furthermore, because the $m^I_{\rm T}$ term in Eq. (\ref{MCA9}) is obtained using non-perturbed eigenstates, the $m^I_{\rm T}$ term is irrelevant to SOI despite being derived from the second-order perturbation of SOI.
In fact, we can obtain the $m_{\rm T}$  using Eqs. (\ref{mtq}) and (\ref{mtspin}) without SOI.
 This is because the approximation in Eq. (\ref{completeness}), which corresponds to neglecting the quantum uncertainty in the angular momentum operators, $\sqrt{\langle L_{\zeta} \rangle ^2-\langle L_{\zeta} ^2 \rangle} \approx 0$.

The above treatment of angular momentum operators results in a classical picture of magnetic anisotropy, namely, the shape magnetic anisotropy (SMA) due to magnetostatic dipole-dipole interactions. St\"{o}hr pointed out that there is a relationship between the magnetic dipole moment derived from the spin-flip term of MCA energies and magnetostatic dipole-dipole interaction (see Appendix B in Ref. \cite{1999Stohr-JMMM}). 
Furthermore, because Eq. (\ref{completeness}) requires the occupied majority-spin and the unoccupied miniroty-spin states in $E_{\rm MCA}^{\sigma \sigma ^{\prime}}$, the magnetic dipole moment (quadrupole moment of spin density) in Eq. (\ref{MCA7})-(\ref{MCA8}) are related to the spin-flip term $E_{\rm MCA}^{\uparrow \downarrow}$ of MCA energies.

\subsection{Intuitive understanding of MCA based on orbital and quadrupole moments}

Based on discussions above, MCA energies can be presented by using the orbital moment anisotropy and the magnetic dipole moment (quadrupole moment of spin density)  as follows:
\begin{equation}
\label{Bruno-WWF}
E_{\rm MCA} \approx \sum _I \frac{1}{4} \xi _I \Delta m_{\rm orb}^I + \frac{21}{8} \sum _I \frac{\xi ^2_I}{\Delta _{\rm exc}} m_{\rm T}^I.
\end{equation}
A difference in the coefficient of the $m_{\rm T}^I$ term by a factor of $\frac{1}{4}$ is noticed in this equation compared to Eq. (\ref{MCA0}). The difference comes from the eigenvalues of the spin operator $\vec{\hat{S}}$ as mentioned  in Eq. (\ref{MCA4}), because Wang, Wu and Freeman described the SOI Hamiltonian by $H^{\rm SO} =\xi \vec{\hat{L}}\cdot\vec{\hat{\sigma}}$ \cite{1993Wang-PRB}, where $\vec{\hat{\sigma}}$ is the Pauli spin matrices (twice of $\vec{\hat{S}}$) and a factor $\frac{1}{2}$ is included in $\xi$. Our formulations of the spin-flip term of MCA energies related to the magnetic dipole moments (quadrupole moments of spin-density) in Eq. (\ref{MCA4})-(\ref{MCA9}) using $H^{\rm SO} =\xi \vec{\hat{L}}\cdot\vec{\hat{S}}$ (thus a factor of $\frac{1}{2}$ is not included in $\xi$) are consistent with the formulation in St\"{o}hr’s paper (see Eq. (27) in Ref. \cite{1999Stohr-JMMM}).  

  Furthermore, it is important to notice that Bruno term (orbital moment) and van der Laan term (quadrupole moment of spin density) \cite{LaanReason} are not sufficient to describe the MCA, because they ignore perturbation terms involving unoccupied majority-spin states\cite{explain}. In spite of the approximation, the Eq. (27) is still important to connect the MCA energy with the local physical quantities observed by the spectroscopic experiments. Following, we add an intuitive picture on MCA to the Bruno term and the van der Laan term. Then, we would like to comment on the difference between Eq. (27) of the present paper and Eq. (28) of Ref. \cite{1998Laan-JPCM}. The Eq. (27) is the MCA energy formulated through the second order perturbation of SOI under the approximation of Eq. (19) and (20). On the other hand, the latter is the energy due to the second-order perturbation derived from Eq. (9) of Ref. \cite{1998Laan-JPCM}, and is not the MCA energy. Thus, to obtain the MCA energy, the energy difference between the perpendicular and in-plane magnetization should be considered. In taking the energy difference, the $E_{LS}$ term in Eq. (28) of Ref. \cite{1998Laan-JPCM} will be canceled out between the perpendicular and in-plane magnetization directions, because the $E_{LS}$ is independent of magnetization direction. Thus, we ignore the $E_{LS}$ term in Eq. (27) of the present paper.

\begin{figure}[tp]
\begin{center}
\includegraphics[height=0.27\textheight,width=0.75\textwidth]{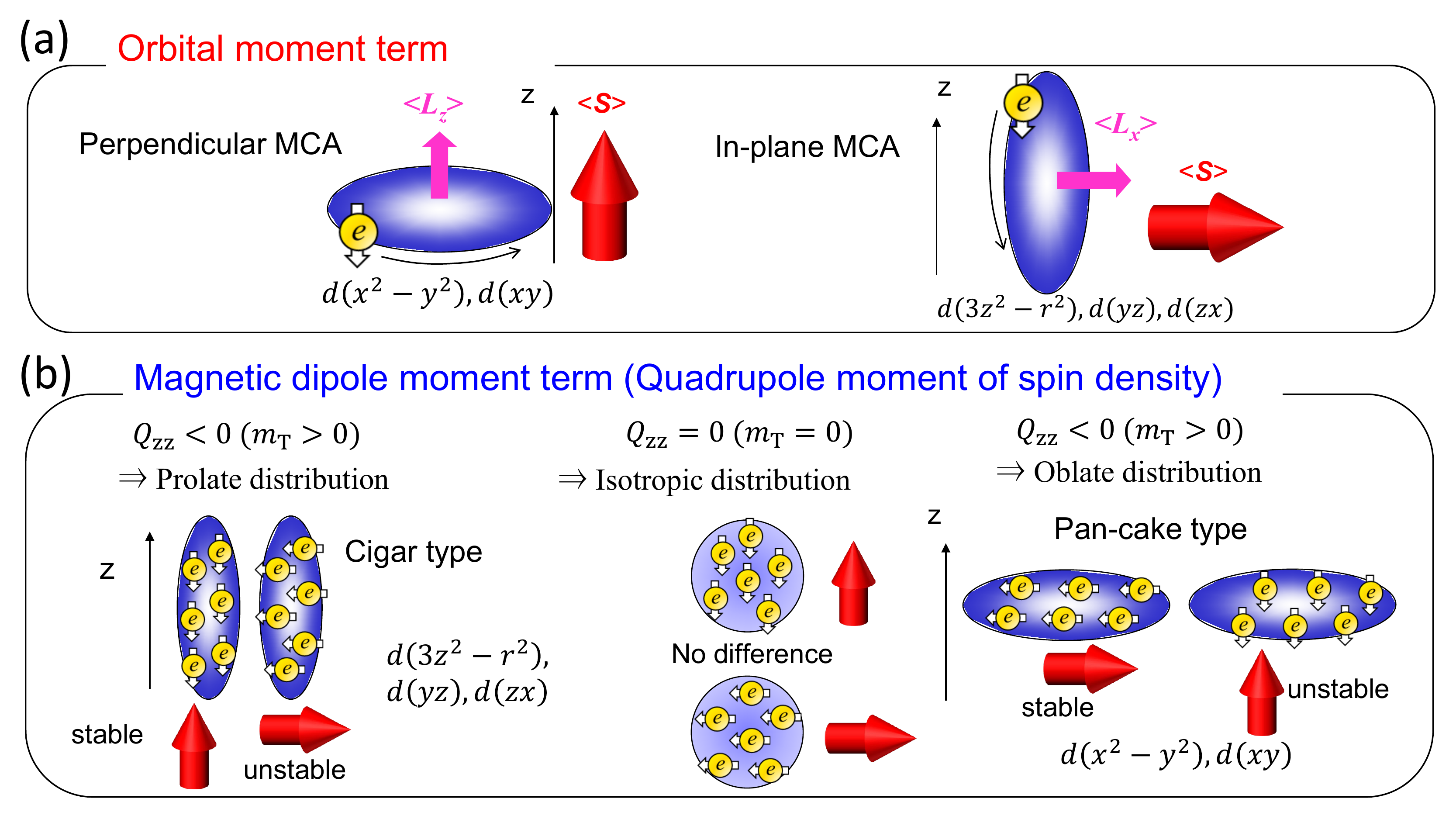}
\end{center}
  \caption{(Color online)Schematic image on effects of the orbital moment anisotropy and quadrupole moment of spin density on MCA. (a)Spin moment can be stabilized along the direction of the larger orbital moment due to SOI, described as the Bruno term in Eq. (\ref{morbdiff}). (b)Spin moment tend to be parallel to the longitudinal direction of the spin-density distribution, characterized by the quadrupole moment ( intra-atomic  magnetic dipole moment) and described as   van der Laan    term in Eq. (\ref{MCA9}).}
\end{figure}

Figure 1 shows schematic images of the orbital and quadrupole moments' contributions to MCA. The contribution of the orbital moment to MCA can be understood from the SOI Hamiltonian in second-order perturbation. Because of the internal product term $\vec{\hat{L}}\cdot\vec{\hat{S}}$ in Eq. (\ref{EPT1}), an orbital moment parallel to the spin moment is more favorable to stabilize the magnetization direction. This implies that the spin moment aligns with a larger orbital moment. Thus, the orbital moment anisotropy $\Delta m^{\rm orb}$ is proportional to the MCA energy. The oblate (pancake-like) distributions of minority-spin electrons in $xy$ plane cause the perpendicular orbital moment (along $z$ axis), contributing to the perpendicular MCA, whereas the prolate (cigar-like) distributions along $z$ axis cause the in-plane orbital moment, contributing to the in-plane MCA (in $xy$ plane).  In contrast, the contribution of the quadrupole moments of spin density (magnetic dipole moments) to MCA implies that the shape of the spin-density distribution directly affects the MCA .

If the quadrupole moments ($Q_{zz}$) of the  spin density defined by Eq. (\ref{qzz}) is zero, the spin-density distribution is spherical and does not contribute to MCA. However, if $Q_{zz}$ is non-zero, the spin moment tends to orient in the longitudinal direction of the spin-density distribution, and the prolate (cigar-type) distribution contributes to the perpendicular MCA, whereas the oblate (pancake-type) distribution contributes to the in-plane MCA. 
Therefore, the contributions to MCA from the two shapes of the spin-density distribution are opposite for the orbital moment and quadrupole moment (magnetic dipole moment). This is consistent with the opposite signs of the matrix elements of $\hat{L}_z$ and $\hat{L}_x$ for the spin-conserving term (orbital moment anisotropy) and the spin-flip term (the quadrupole moment of spin density).
 Therefore, the MCA of magnetic materials with TMs can be determined from the competition between the orbital moment anisotropy and the anisotropy of the spin-density distribution (the quadrupole moment of spin-density).

Recently, Suzuki {\it et al.} \cite{2019Suzuki-PLA} formulated  the expectation of the magnetic dipole operator by using wavefunctions with first-order perturbative corrections of SOI, and directly describe the spin-flip term of perturbative MCA energies with the measurable physical parameters related to the magnetic dipole moment. The formulation does not require the approximation and assumption such as Eqs. (\ref{exchange_splitting}) and (\ref{completeness}), and can apply magnetic materials with small exchange splitting.
However, an intuitive picture between the MCA and SOI corrections of the magnetic dipole term is still unclear; thus, that is a future work. 

\begin{table}[tp]
\centering
\caption{\label{exp_Ni} The orbital magnetic moment of Ni in Ni/Cu multilayer on BaTiO$_3$ measured by XMCD for Ni $L$ edges in the normal incident (NI) and the grazing incident (GI) geometries, with and without electric field ($E$) . The calculated orbital magnetic moments of fcc Ni for the perpendicular ($\theta =0$) and in-plane ($\theta=\pi/2$) magnetization configurations with and without tensile distortion (2 \%), corresponding to with and without electric field ($E$)  in the experiment.}
\footnotesize
\begin{tabular}{c|c|c}
\br
 Experiment & $E=0$ kV/cm  & $E=8$ kV/cm \\
\cline{1-1}\cline{2-2}\cline{3-3} 
$m_{\rm orb}$ (NI) &  0.06  $\mu _{\rm B}$&  0.04  $\mu _{\rm B}$\\
$m_{\rm orb}$ (GI) & 0.04  $\mu _{\rm B}$&  0.05  $\mu _{\rm B}$\\
\cline{1-3}
Calculation  & 2 \% tensile strain &  without strain  \\
\cline{1-1}\cline{2-2}\cline{3-3} 
$m_{\rm orb}$ ($\theta=0$) &  0.0542 $\mu _{\rm B}$&  0.0506  $\mu _{\rm B}$\\
$m_{\rm orb}$ ($\theta=\pi/2$) & 0.0505  $\mu _{\rm B}$&  0.0504  $\mu _{\rm B}$\\
\br

\end{tabular}
\end{table}

\section{Fcc Ni with  in-plane distortions}

The coupling between MCA and lattice distortion is a fundamental issue in magnetism. The lattice distortion changes the symmetry of a crystal and produces different electronic structures affecting magnetization directions.  Therefore, magnetic properties and electronic structures strongly couple with lattice structure and symmetry, known as the inverse magnetostriction (magneto-elastic) effect. For example, fcc Ni does not have MCA in the cubic structure. However, the tensile and compressive tetragonal lattice distortions with respect to the (001) plane cause the perpendicular and in-plane MCAs, respectively.

\begin{figure}[tp]
\begin{center}
\includegraphics[height=0.27\textheight,width=1.0\textwidth]{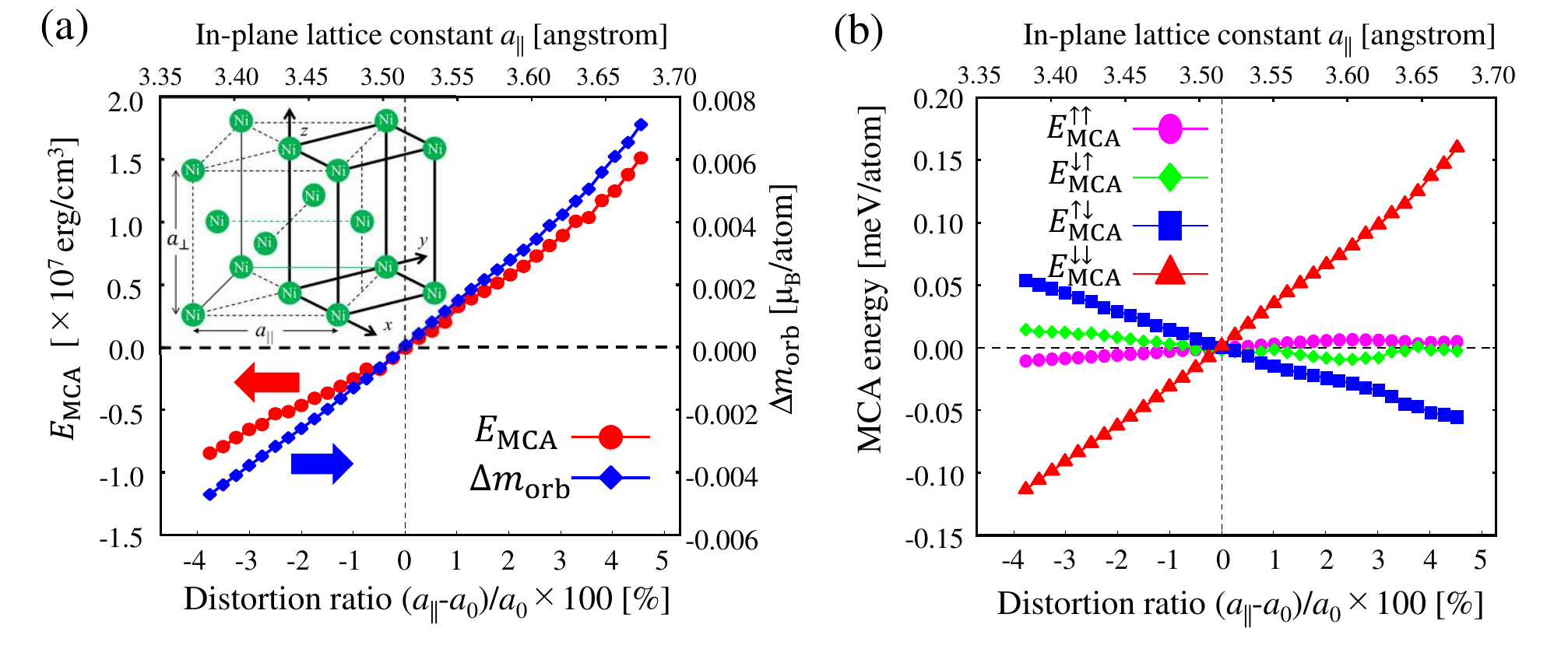}
  \caption{(Color online)(a) DFT calculation results of MCA energies and the orbital moment anisotropies of fcc Ni as a function of the in-plane lattice constant $a_{\parallel}$ (tetragonal distortion is defined as ($a_{\parallel}-a_0$)/$a_0\times100$), where $a_0$ = 3.524 \AA. The crystal structure of fcc Ni for the conventional unit cell and the unit cell with 45$^{\circ}$ in-plane rotation is also shown in the inset. (b) Spin-resolved MCA energies  in the second-order perturbation of SOI  from Eq. (\ref{MCA2}) for fcc Ni as a function of $a_{\parallel}$.}
\end{center}
\end{figure}

Recently, the magnetic properties of Ni/Cu multilayers on ferroelectric BaTiO$_3$ substrate were controlled by the mechanical strain  through the piezo electric effect . This strain is induced by an applied electric field $E$ to BaTiO$_3$, switching the magnetization from the perpendicular to in-plane easy axis by tuning the lattice distortion with $E$ \cite{2019Okabayashi-npjQM}. In this experiment, without $E$, the Ni layer exhibited a tensile strain of 2 \% through the sandwiched Cu layers, and the Ni layer showed a perpendicular MCA. With $E$, the change in the domain structures reduces the lattice constant of BaTiO$_3$ and releases the tensile strain in the Ni layers, resulting in magnetization along the in-plane easy axis. In addition, a strain-induced change in the orbital magnetic moments of Ni was observed by XAS and XMCD measurements. The spectra for NiCu/BaTiO$_3$ were measured for Ni $L$ edges in the normal incident (NI) and grazing incident (GI) geometries, with and without $E$ (8 kV/cm).

Table 2 shows the observed orbital moments of Ni in the Ni/Cu multilayers on BaTiO$_3$ by XMCD measurement.
The orbital moments in the NI geometry correspond to those with the magnetization normal to the (001) plane, while the orbital moments in the GI geometry include half of the in-plane components owing to the grazing angle (60$^{\circ}$) from the sample surface normal ($\cos 60 ^{\circ}=1/2$ ). 
The values of the orbital moment anisotropy of Ni between the NI and GI geometries were estimated to be 0.02 $\mu_{\rm B}$ ( for $E$ = 0 and tensile strain = 2\%) and 0.01 $\mu_{\rm B}$ ( for $E$ = 8 kV/cm and tensile strain = 0).  
These results indicate that an applied $E$ modulates the orbital moments and their anisotropies, resulting in changes in MCA owing to lattice distortion.

To examine the relationship between MCA and lattice distortion in detail, we performed second-order perturbation calculations for the MCA of fcc Ni with tetragonal distortion in the (001) plane. Details of DFT calculations are presented in Ref. \cite{2019Okabayashi-npjQM} .
Figure 2(a) shows the MCA energy and orbital moment anisotropy of fcc Ni as a function of the distortion ratio $(a_{\parallel}-a_0)/a_0\times 100$, where $a_{\parallel}$ is the in-plane lattice parameter, and $a_0=3.524$ \AA~ is the optimized lattice constant of the cubic fcc Ni in DFT calculations. The inset of Fig. 2(a) shows the crystal structure and unit cell of distorted fcc Ni. The out-of-plane lattice parameter $a_{\perp}$ is fully relaxed for each $a_{\parallel}$.

As shown in Fig. 2(a), the MCA energy increases with an increasing distortion ratio (i.e., tetragonal tensile strain), which is consistent with the experimental results.
Because the orbital moment anisotropy $\Delta m_{\rm orb}$ monotonically increases with an increasing distortion ratio, the change in the perpendicular MCA  can be attributed to the change in the orbital moment anisotropy given by Eq. (\ref{morbdiff}). Figure 2(b) shows the dependence of the spin-resolved MCA energies in second-order perturbation from Eq. (\ref{MCA2}) on the tetragonal strain. The spin-conserving term $E_{\rm MCA}^{\downarrow \downarrow}$ has the largest dependence on the distortion ratio  and is the main contributing factor to MCA. In contrast, for the spin-flip term $E_{\rm MCA}^{\uparrow \downarrow}$, the magnitude is smaller than the spin-conserving term, and $E_{\rm MCA}^{\uparrow \downarrow}$ has the opposite dependence on the distortion ratio compared to MCA energies.

\begin{figure}[tp]
\begin{center}
\includegraphics[height=0.25\textheight,width=1.0\textwidth]{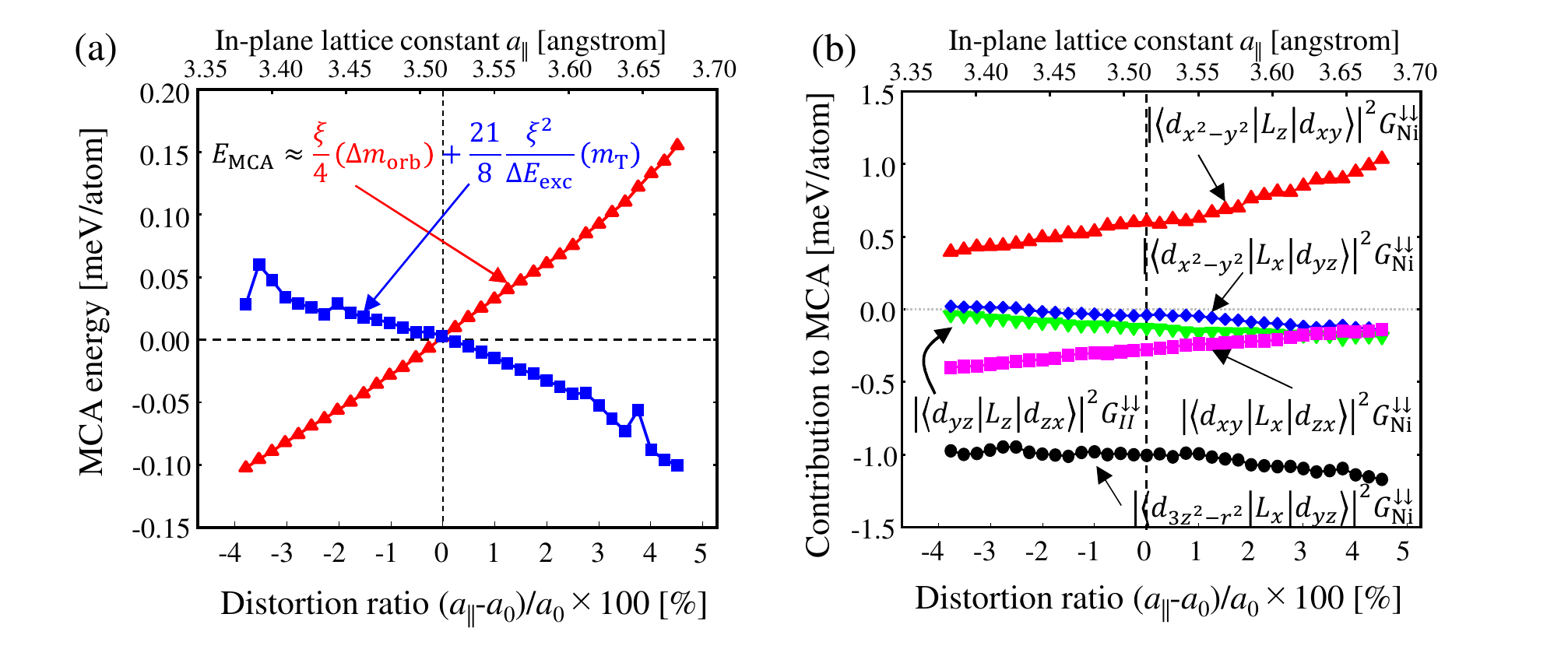}
\end{center}
  \caption{(Color online)(a) Bruno term (orbital moment anisotropy) expressed by Eq. (\ref{morbdiff}) and   van der Laan    term (magnetic dipole moment) expressed by Eq. (\ref{MCA9}) of fcc Ni as a function of $a_{\parallel}$. (b) The $d$ orbital contribution to the MCA energies of tetragonally distorted fcc-Ni as a function of the in-plane lattice constant $a_{\parallel}$, where $G_{\rm Ni}^{\downarrow \downarrow}$ is the joint local density of states expressed by Eq. (\ref{JLDOS}).}
\end{figure}

To validate Eq. (\ref{Bruno-WWF}), we plotted the Bruno term from Eq. (\ref{morbdiff}) in Fig. 3(a) and   van der Laan    term from Eq. (\ref{MCA9}) as a function of the distortion ratio, where $\Delta E_{\rm exc}$ was set to 1 eV. By comparing Fig. 2(b) and Fig. 3(a), we found that the strain dependence of Bruno and   van der Laan    terms (on the distortion ratio) agrees with that of the spin-conserving term $E_{\rm MCA}^{\downarrow \downarrow}$ and the spin-flip term $E_{\rm MCA}^{\uparrow \downarrow}$, respectively, indicating that the orbital moment anisotropy and the magnetic dipole (the quadrupole moment of spin density) successfully describe the MCA of distorted fcc Ni. In this case, the orbital moment anisotropy is more responsible for the MCA of fcc Ni with tetragonal distortion than the anisotropy of the quadrupole moment (the magnetic dipole term).

To further  understand the MCA of fcc Ni, in Fig. 3(b), we show the square of the matrix elements of the spin-conserving terms corresponding to $L_z$ and $L_x$ for each $d$ orbital  as a function of the distortion ratio. 
The matrix elements $|\langle d_{xy}|L_z|d_{x^2-y^2} \rangle|^2G_{\rm Ni}^{\downarrow\downarrow}$  in Eq. (\ref{MCA3})   show the main contribution to the perpendicular MCA, and their lattice distortion dependence is similar to that of $E_{\rm MCA}$, i.e., it increases with increasing distortion ratio,  where the positive (negative) matrix elements indicate the contribution to the perpendicular (in-plane) MCA. 
We confirm that the number of minority-spin electrons in $d_{x^2-y^2}$ increases owing to tensile distortion, whereas that in $d_{xy}$ changes negligibly. 
This change in electron distribution in $d_{x^2-y^2}$ under tensile distortion enhances the matrix element $|\langle d_{xy}|L_z|d_{x^2-y^2} \rangle|^2G_{\rm Ni}^{\downarrow\downarrow}$ and the orbital moment along the perpendicular direction, leading to the origin of the perpendicular MCA of fcc Ni under tensile distortion.

\begin{figure}[tp]
\begin{center}
\includegraphics[height=0.5\textheight,width=1.0\textwidth]{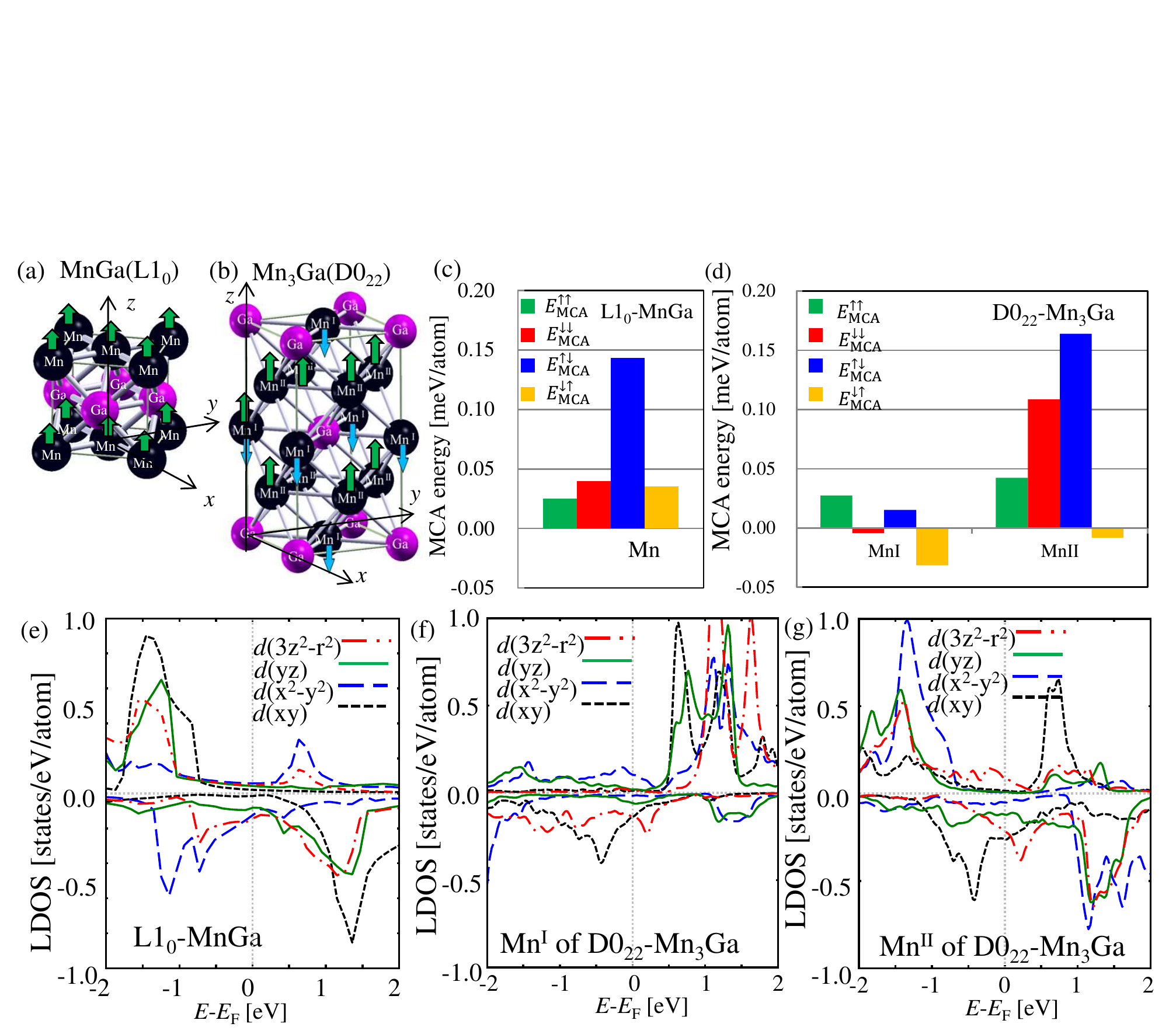}
\end{center}
  \caption{(Color online)Schematics of crystal structures and spin configurations of (a)L1$_0$-MnGa and (b)D0$_{22}$-Mn$_3$Ga. Spin-resolved MCA energies in the second-order perturbation of the SOI for (c) Mn in L1$_0$-MnGa and (b) Mn$^{\rm I}$ and Mn$^{\rm II}$ in D0$_{22}$-Mn$_3$Ga.   Local density of states (LDOS) of each $d$ orbital as a function of energy relative to the Fermi energy $E_{\rm F}$ for (e) Mn in L1$_0$-MnGa, (f) Mn$^{\rm I}$ and (g) Mn$^{\rm II}$ in D0$_{22}$-Mn$_3$Ga.   }
\end{figure}

\section{Mn-Ga alloys}

In Section 3, we introduced a bulk system with a perpendicular MCA originating from the orbital moment anisotropy. We now discuss a perpendicular MCA arising from the quadrupole moment of the spin density.
Recently, XMCD and XMLD measurements were performed to detect the orbital and quadrupole moments in Mn-Ga binary alloys to clarify the origin of the perpendicular MCA of Mn$_{3-\delta}$Ga alloys \cite{2020Okabayashi-SciRep}.
 Mn$_{3-\delta}$Ga is one of the candidates that could overcome some of the problems in spintronics devices by reducing the energy consumption during magnetization reversal and enhancing the thermal stability due to their strong perpendicular MCA with the  ferrimagnetic property.

\begin{table}[tp]
\centering
\caption{\label{calc_MnGa}Calculated MCA energy $E_{\rm MCA}$, the spin moment $m_{\rm spin}$, the orbital moment anisotropy $\Delta m_{\rm orb}$, and the magnetic dipole moment $m_{\rm T}$ of L1$_0$-MnGa and D0$_{22}$-Mn$_3$Ga with the experimental lattice constant \cite{2016Suzuki-SciRep,2007Balke-APL}. }
\footnotesize
\begin{tabular}{c|c|c}
\br
 Calculation & L1$_0$-MnGa & D0$_{22}$-Mn$_3$Ga  \\
\cline{1-1}\cline{2-2}\cline{3-3} 
$E_{\rm MCA}$ &  1.463  MJ/m$^3$& 1.856  MJ/m$^3$ \\
$m_{\rm spin}$ & 2.667 $\mu _{\rm B}$ & -3.177  $\mu _{\rm B}$ (Mn$^{\rm I}$) 2.546  $\mu _{\rm B}$ (Mn$^{\rm I\hspace{-1pt}I}$) \\
$\Delta m_{\rm orb}$ &  0.0014  $\mu _{\rm B}$&  -0.0030  $\mu _{\rm B}$ (Mn$^{\rm I}$) 0.0064  $\mu _{\rm B}$ (Mn$^{\rm I\hspace{-1pt}I}$) \\
$m_{\rm T}$ & 0.0493  $\mu _{\rm B}$&  0.0180  $\mu _{\rm B}$ (Mn$^{\rm I}$) 0.0801  $\mu _{\rm B}$ (Mn$^{\rm I\hspace{-1pt}I}$) \\

\br

\end{tabular}
\end{table}

Schematics of the crystal structures of L1$_0$-MnGa($\delta=2$) and D0$_{22}$-Mn$_3$Ga($\delta$=0) are shown in Figs. 4(a) and (b), respectively.
 In Mn$_3$Ga, the Mn$^{\rm I}$ is located at the same (001) plane with Ga having negative spin moment, while Mn$^{\rm I\hspace{-1pt}I}$ is located between the two (001) Mn$^{\rm I}$-Ga planes having positive spin moment.  
  Detecting the element specific quadrupole tensor $Q_{zz}$ is possible with the XLMD measurements and the sum rule in the Mn $L$  edges. According to the XMCD and XMLD measurements, MnGa and Mn$_3$Ga had relatively large magnetic dipole moments $m_{\rm T}$ of Mn (of the order of 0.01 $\mu_{\rm B}$) while the orbital moment anisotropies $\Delta m_{\rm orb}$ were negligibly small (less than 0.01) despite the perpendicular MCA of MnGa and Mn$_3$Ga.

\begin{figure}[tp]
\begin{center}
\includegraphics[height=0.27\textheight,width=1.0\textwidth]{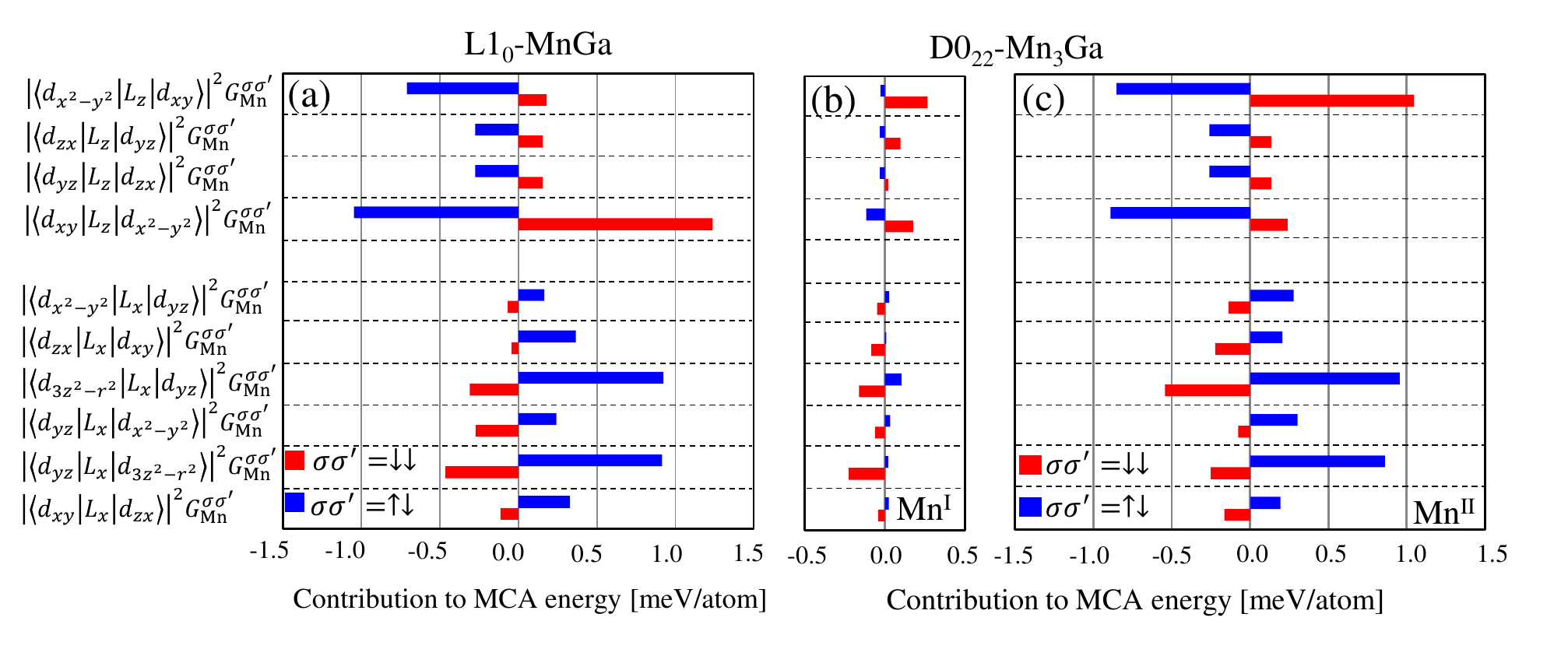}
\end{center}
  \caption{(Color online)Matrix elements of the second order perturbation of SOI between $d$-orbitals of Mn atoms in L1$_0$ MnGa and D0$_{22}$-Mn$_3$Ga for the spin-conserving ($\downarrow\downarrow$) and the spin-flip ($\uparrow\downarrow$) processes, where $G_{\rm Mn}^{\downarrow \downarrow}$ is the joint local density of states expressed by Eq. (\ref{JLDOS}). Positive and negative values indicate the contribution of the matrix elements to the perpendicular and the in-plane MCA, respectively.  }
\end{figure}

 Table 3 shows the calculated spin moments, orbital moment anisotropies $\Delta m_{\rm orb}$, and magnetic dipole moments $m_{\rm T}$ of Mn in L1$_0$-MnGa and D0$_{22}$-Mn$_3$Ga with the experimental lattice constant shown in Ref. \cite{2016Suzuki-SciRep} and Ref. \cite{2007Balke-APL}, respectively. The details of DFT calculations are described in Ref. \cite{2020Okabayashi-SciRep}. The calculated magnetic dipole moments are larger than the orbital moment anisotropies both in MnGa and Mn$_3$Ga, demonstrating consistency with the experimental results. This indicates that the magnetic dipole moment related to the cigar-type quadrupole moment of the spin density plays an important role in the perpendicular MCA of these alloys. 
Figures 4(c) and (d) show the spin-resolved MCA energies in the second-order perturbation of the SOI for MnGa and Mn$_3$Ga.
We found  that the spin-flip term ($E^{\uparrow\downarrow}_{\rm MCA}$) of Mn in MnGa and Mn$^{\rm II}$ in Mn$_3$Ga are the main contributors to MCA energies.
Since Mn$_3$Ga is a ferrimagnet, Mn$^{\rm I}$ and Mn$^{\rm II}$ have mutually opposite spin directions, where the contribution of Mn$^{\rm I}$ to MCA is smaller than that of Mn$^{\rm II}$.
As discussed in Section 2, the spin-flip term $E^{\uparrow\downarrow}_{\rm MCA}$ of the MCA energy  is described by the magnetic dipole moment, indicating that the origin of the perpendicular MCA can be attributed to the cigar-type quadrupole moment of the spin density rather than the orbital moment anisotropies.
  In Fig. 4(e)-(g), we show the local density of states (LDOS) of L1$_0$-MnGa and D0$_{22}$-Mn$_3$Ga by the DFT calculation. In the LDOS of the Mn$^{\rm I}$ and Mn$^{\rm II}$ sites, all orbital states were split through exchange interaction. However, exchange splitting was incomplete where complete spin splitting was required, in the Bruno formula, which enabled the transitions by spin mixing between occupied spin-up and unoccupied spin-down states.

To further understand how atomic orbitals contribute to the perpendicular MCA in L1$_0$ MnGa and D0$_{22}$-Mn$_3$Ga, we shows in Fig. 5 the square of the matrix elements of  $L_z$ and $L_x$ for each $d$ orbital of Mn in the spin-conserving and spin-flip terms.
As shown in Fig. 5(a), the matrix elements that contribute positively to the MCA energy of L1$_0$-MnGa (perpendicular MCA) are $|\langle d_{yz}|L_x|d_{3z^2-r^2} \rangle|^2G_{\rm Mn}^{\uparrow\downarrow}$ and $|\langle d_{3z^2-r^2}|L_x| d_{yz} \rangle|^2G_{\rm Mn}^{\uparrow\downarrow}$, both of which are spin-flip matrix elements. Although the matrix element $|\langle d_{xy}|L_z|d_{x^2-y^2} \rangle|^2G_{\rm Mn}^{\downarrow\downarrow}$ shows a large positive value, 
its spin-flip term $|\langle d_{xy}|L_z|d_{x^2-y^2} \rangle|^2G_{\rm Mn}^{\uparrow\downarrow}$ shows a large negative value, leading to a small contribution of $d_{xy}$ and $d_{x^2-y^2}$ orbitals to the perpendicular MCA because each term is canceled out. 
For D0$_{22}$-Mn$_3$Ga, the matrix elements of Mn$^{\rm I}$ are much smaller than those of Mn$^{\rm II}$. 
Furthermore, the spin-flip matrix elements of $L_x$ between $d_{3z^2-r^2}$ and $d_{yz}$ of Mn$^{\rm II}$ had large positive values, indicating the origin of the perpendicular MCA. Therefore, the perpendicular MCA of L1$_0$ MnGa and D0$_{22}$-Mn$_3$Ga originates from the cigar-type distribution of spin density, especially because of $d_{3z^2-r^2}$ and $d_{yz}$ orbitals, where the spin moment of Mn tends to be oriented in the longitudinal (perpendicular) direction of the spin-density distributions.

\begin{figure}[tp]
\begin{center}
\includegraphics[height=0.27\textheight,width=1.0\textwidth]{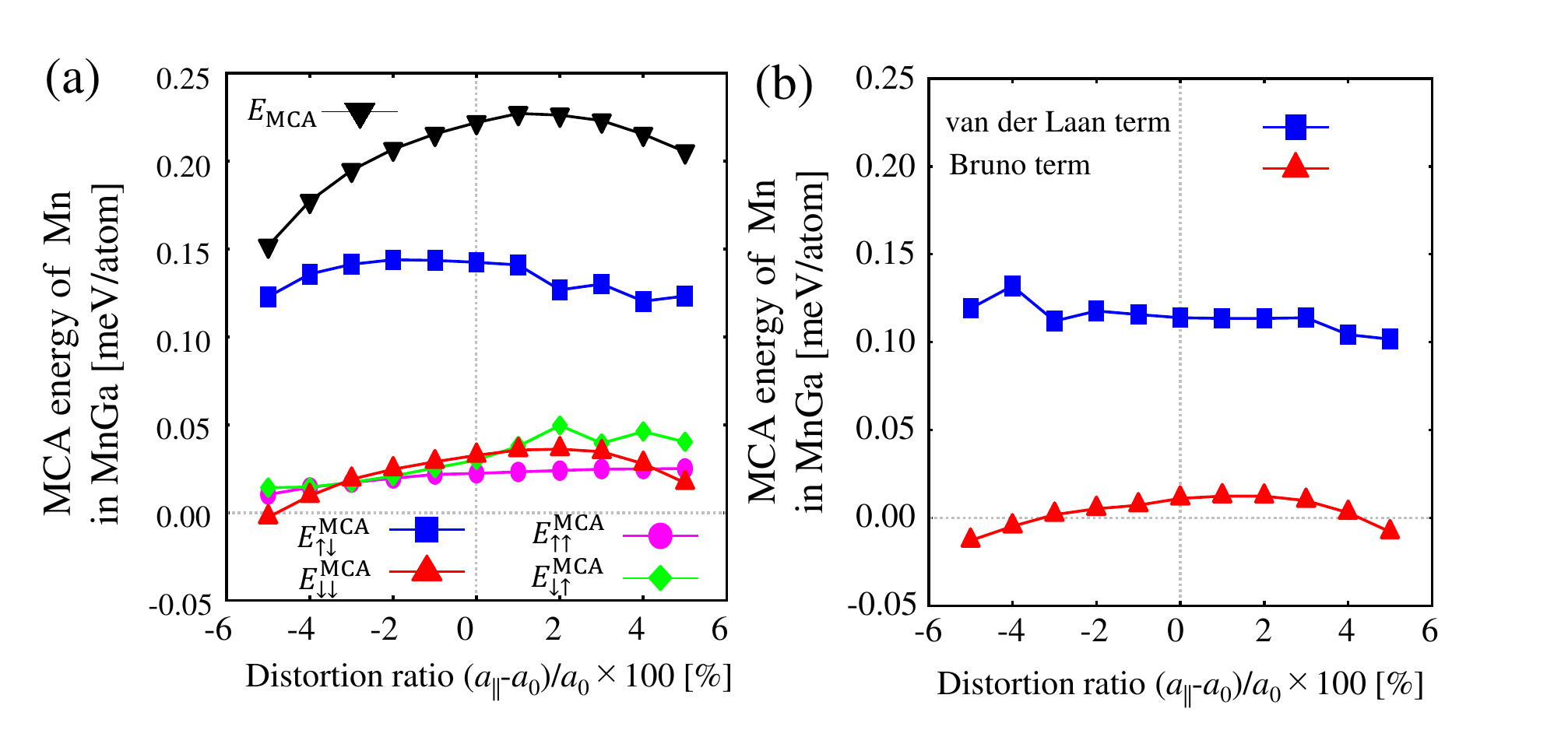}
\end{center}
  \caption{(Color online )(a) Spin-resolved MCA energies in the second-order perturbation of SOI for L1$_0$-MnGa shown in Eq. (\ref{MCA2}) as a function of $a_{\parallel}$. (b) Bruno term shown in Eq. (\ref{morbdiff}) and   van der Laan   term shown in Eq. (\ref{MCA9}) of L1$_0$-MnGa as a function of $a_{\parallel}$.}
\end{figure}

Because the dependence of MCA on lattice distortion is an important physical aspect, we investigated the change in MCA energies with tetragonal distortion.
Figure 6(a) shows MCA energies and MCA contributions of each spin transition process in the second-order perturbation for Mn atom in L1$_0$-MnGa as a function of the in-plane distortion ratio. 
We find that the spin-flip term $E_{\rm MCA}^{\uparrow \downarrow}$ has the largest contribution to the perpendicular MCA. Whereas its dependence on the in-plane distortion ratio is weaker than that on the total MCA energy.
In contrast, the spin-conserving term $E_{\rm MCA}^{\downarrow \downarrow}$ exhibits a stronger dependence on in-plane distortion ratio and connects MCA energies to lattice distortion,  which is consistent with the results for fcc Ni in Fig. 3.

The above  behavior is also confirmed in Fig. 6(b), where Bruno term in Eq. (\ref{morbdiff}) and   van der Laan   term in Eq. (\ref{MCA9}) for Mn in L1$_0$-MnGa are plotted as a function of the distortion ratio.
Here, we used $\Delta E_{\rm exc}=2$ eV. 
These results indicate that the orbital moment anisotropy is more sensitive to lattice distortion than the magnetic dipole moment (quadrupole moment of spin density), which can be understood in terms of an orbital-striction or an orbital-elastic effect as is discussed in Ref. \cite{2019Okabayashi-npjQM}.

\begin{table}[tp]
\centering
\caption{\label{CoPd} The spin moment $m_{\rm spin}$, the orbital moment anisotropy $\Delta m_{\rm orb}$, and the magnetic dipole $m_{\rm T}$ of interface Co and Pd in Co(4ML)/Pd(8ML)(111) multilayer obtained from the XMCD measurement and DFT calculations}
\footnotesize
\begin{tabular}{c|c|c|c|c}
\br
  & \multicolumn{2}{c|}{Co}       &       \multicolumn{2}{|c}{Pd}             \\
\cline{1-5}
 $\mu _{\rm B}$ & XMCD  & DFT &XMCD & DFT  \\
\cline{1-1}\cline{2-3}\cline{4-5} 
$m_{\rm spin}$   &  1.82  & 1.87  & 0.25 & 0.31 \\
$\Delta m_{\rm orb}$   &  0.03  &  0.032      & 0.00  & -0.001  \\
$m_{\rm T}$              & 0.0014  &  -0.0137 & 0.0014 & 0.0106 \\
\br
\end{tabular}
\end{table}

\section{Co/Pd(111) multilayers}
So far, we have discussed MCA in bulk materials. In this section, we discuss  MCA for interface systems. 
Among various interface combinations, the interaction between 3$d$ TMs and 4$d$ or 5$d$ TMs is considered for a study of the interface-driven MCA because of the large spin moment of 3$d$ TMs and the strong SOI in 4$d$ and 5$d$ TMs.

Ultrathin Co/Pd(111) multilayers are typical artificial nanosystems exhibiting an interface perpendicular MCA. The development of synthesized thin films with perpendicular-magnetization has led researchers to expect ultrahigh density recording media.
Furthermore, using Co/Pd interfaces and multilayers, researchers have demonstrated the photo-induced precession of magnetization \cite{2010Boeglin-Nature,2013Yamamoto-IEEETransMag}, creation of skyrmions using the interfacial Dzyaloshinskii-Moriya interaction \cite{2017Pollard-NC}, and magnetization reversal using the spin-orbit torque \cite{2013Jamali-PRL}. 
Despite much interest in Co/Pd interfaces, the mechanism of the perpendicular MCA and the role of Co and Pd sites were not fully understood.
To clarify the origin of the perpendicular MCA of Co/Pd(111), element-specific XMCD measurements for both Co and Pd were performed on ultrathin films of Co/Pd(111) multilayers  \cite{2018Okabayashi-SciRep}.

\begin{figure}[tp]
\begin{center}
\includegraphics[height=0.26\textheight,width=0.9\textwidth]{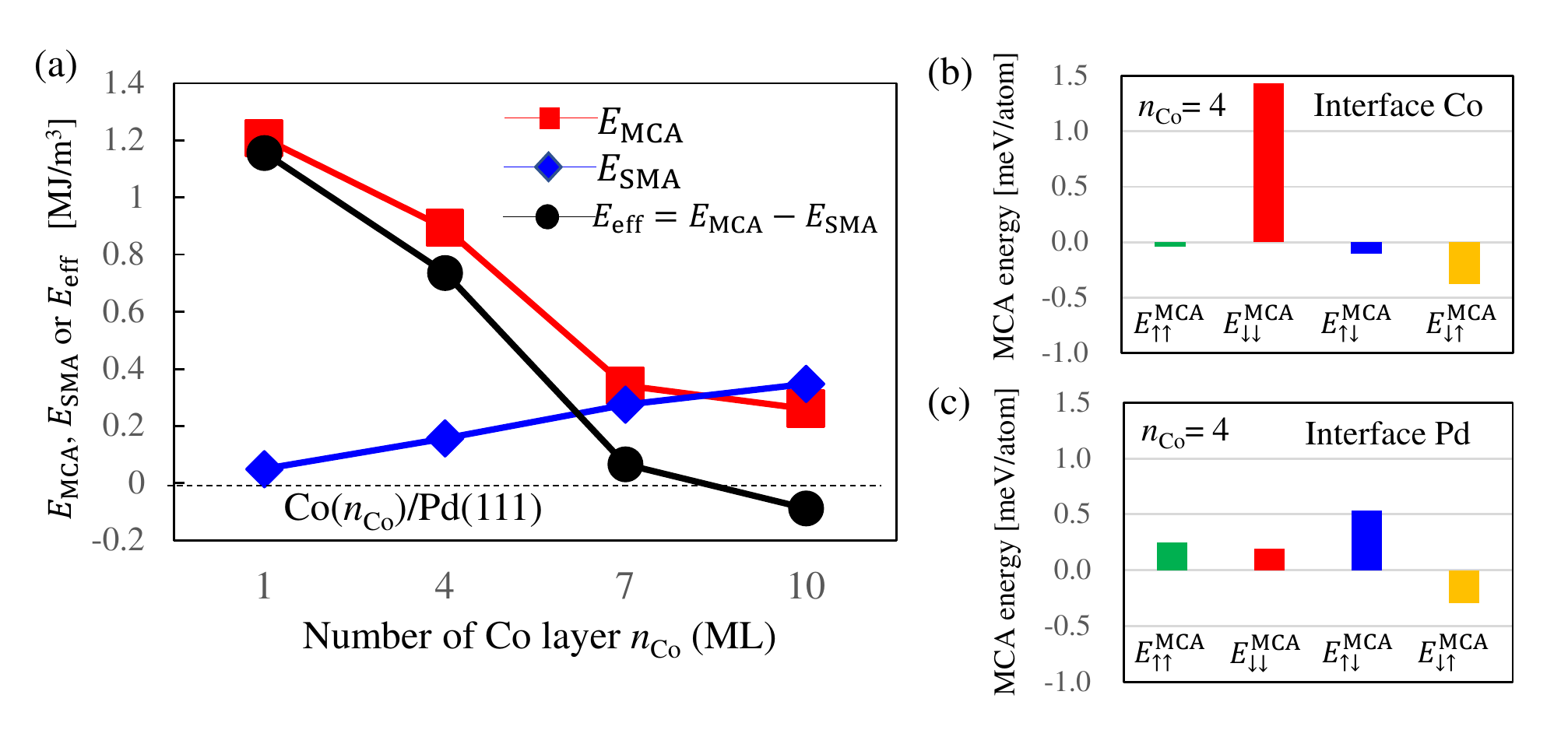}
\end{center}
  \caption{(Color online)The dependence of the MCA energy ($E_{\rm MCA}$), the SMA energy ($E_{\rm SMA}$), and the effective magnetic anisotropy energy ($E_{\rm eff}=E_{\rm MCA}-E_{\rm SMA}$) on the number of Co layer ($n_{\rm Co}$) for Co($n_{\rm Co}$)/Pd(8 ML)(111) multilayer. (b)Spin-resolved MCA energies in the second-order perturbation of SOI from Eq. (\ref{MCA2}) for interface Co and (c)Pd of Co(4 ML)/Pd(111).}
\end{figure}

In the experiments, ultra-thin multilayered samples of [Co (4 ML)/Pd (8 ML)(111)]$_5$ with an out-of-plane easy axis were fabricated. A thicker Co monolayer (ML) case exhibit in-plane MCA because of large SMA. 
Then, the spectra of these samples were measured for the Pd $M_{2,3}$ and Co $L_{2,3}$ edges in the NI and GI geometries. 
Table 4 presents the experimental values of $m_{\rm spin}$, $\Delta m_{\rm orb}$, and $m_{\rm T}$ for the sample having out-of-plane magnetization together with those estimated from DFT calculations.

From Table  4, a positive $\Delta m_{\rm orb}$ for Co sites and its negligible value for Pd sites were observed in XMCD, indicating that the Bruno term (\ref{morbdiff}) in the Co sites is the origin of the perpendicular MCA of Co/Pd(111). 
In contrast, a finite $m_{\rm T}$ was observed at approximately 0.01 $\mu _{\rm B}$ for both Co and Pd sites. Because the SOI constant of Pd atoms $\xi _{\rm Pd}$ is three times that of Co atoms $\xi _{\rm Co}$, the contribution of   van der Laan    term (\ref{MCA9}) to the perpendicular MCA for Pd atoms is three times that of Co atoms. Furthermore, the observed $\Delta m_{\rm orb}$ of Co was negligibly small for the sample with thicker Co layers having in-plane magnetization, indicating that $\Delta m_{\rm orb}$ of the interface Co is responsible for the perpendicular MCA.

\begin{figure}[tp]
\begin{center}
\includegraphics[height=0.25\textheight,width=0.75\textwidth]{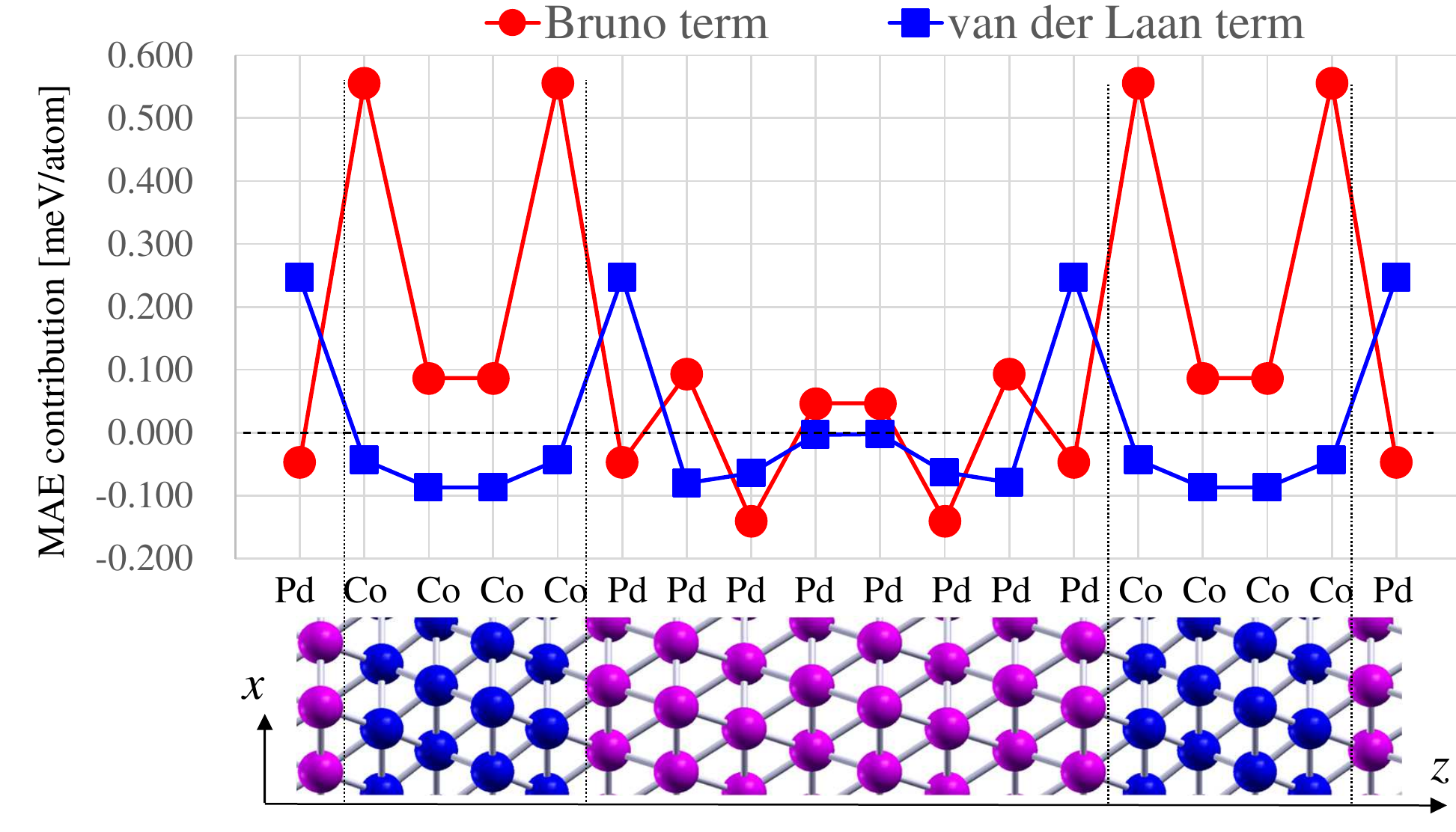}
\end{center}
  \caption{(Color online)Bruno term shown in Eq. (\ref{morbdiff}) and   van der Laan     term shown in Eq. (\ref{MCA9}) at each atomic site of Co and Pd in Co(4ML)/Pd(8ML)(111) multilayer. The atomic structure of Co(4ML)/Pd(8ML)(111) is shown below the graph, where each atomic position along $z$ axis corresponds to the points in the graph.}
\end{figure}

Fig. 7(a) shows $E_{\rm MCA}$, the SMA energy ($E_{\rm SMA}=\frac{1}{2\mu _0}M_{\rm s}^2$), and the effective magnetic anisotropy energy $E_{\rm eff}=E_{\rm MCA}-E_{\rm SMA}$ of Co($n_{\rm Co}$ ML)/Pd(8 ML)(111), where $M_{\rm s}$ is the total magnetization of Co/Pd(111). The in-plane lattice constants and out-of-plane atomic distances of the Co($n_{\rm Co}$)/Pd(111) supercells were fully optimized for each the number of Co layers ($n_{\rm Co}$) by DFT calculations. Other details of DFT calculations are presented in Ref.  \cite{2018Okabayashi-SciRep}. $E_{\rm MCA}$ (red line points) decreases with an increase in $n_{\rm Co}$, indicating that interface Co has the main contribution to MCA.
Because the SMA energy of Co/Pd(111) (blue line points) increases with increasing $n_{\rm Co}$, Co/Pd(111) has a negative $E_{\rm eff}$ around $n_{\rm Co}=6\sim7$ and prefers in-plane magnetization, which is consistent with the experimental result.

To confirm these results, we performed the second-order perturbation analyses of Co/Pd(111) multilayers.
Figures 7(b) and  (c) show the spin-resolved MCA energies in the second-order perturbation of the SOI for the interface Co and Pd atoms in Co(4ML)/Pd(111).
We find that for the interface Co, the spin-conserving term $E^{\downarrow\downarrow}_{\rm MCA}$ is the main contributor  to the perpendicular MCA. Whereas for the interface Pd, the spin-flip term $E^{\uparrow\downarrow}_{\rm MCA}$ shows the main contribution. Because $E^{\downarrow\downarrow}_{\rm MCA}$ and $E^{\uparrow\downarrow}_{\rm MCA}$ are related to $\Delta m_{\rm orb}$ and $m_{\rm T}$, these results are also consistent with the experimental results in Table 4.

To examine the layer-by-layer contributions to MCA energies of Co/Pd(111), we show in Fig. 8 Bruno and   van der Laan     terms estimated from $\Delta m_{\rm orb}$ and $m_{\rm T}$ at each atomic site of Co(4ML)/Pd(8ML)(111). 
To obtain   van der Laan    term, we use $\Delta _{\rm exc}=4$ eV.  It is noted that the value of exchange splitting $\Delta _{\rm exc}=4$ eV is too large for nonmagnetic element Pd. However, the $\Delta _{\rm exc}$ is defined by Eq. (\ref{exchange_splitting}) indicating the energy range of eigenstate $\epsilon _{\vec{k},n}$. Because the valence states of Pd in Co/Pd(111) shows broad  energy range corresponding to $4\sim5$ eV due to delocalized character of 4$d$ element (see Fig. 5(d) of Ref. \cite{2018Okabayashi-SciRep}), we use 4 eV for $\Delta _{\rm exc}$.   
As shown in Fig. 8, the Bruno term increases only at the interface Co sites, whereas   van der Laan     term at the Pd sites. This indicates that an interface-driven perpendicular MCA originates from both the orbital moment anisotropy and cigar-type quadrupole moment of spin density (corresponding to positive magnetic dipole moment). In Co/Pd(111), owing to the hybridization between the Co 3$d$ and Pd 4$d$ orbitals, there is large proximity  that induces SOI in the Co states and spin polarization in the Pd states, which leads to a large induced spin moment in Pd of approximately 0.3 $\mu _{\rm B}$.

Moreover, the strong  proximity effect in the Co/Pd(111) multilayer leads to the orbital moment anisotropy in Co site $\Delta m_{\rm orb}$ and the magnetic dipole moment in Pd site $m_{\rm T}$ corresponding to the quadrupole moment of spin density.
From the LDOS of interfacial Co in the Co/Pd(111) multilayer, the Co 3$d_{xy}$, 3$d_{x^2-y^2}$, 3$d_{yz}$, and 3$d_{zx}$ orbitals contribute to the perpendicular MCA because of being dominant at the Fermi level in the minority-spin states \cite{2018Okabayashi-SciRep}. In addition, these states provide large spin-conserving matrix elements of the interface Co,  such as $|\langle d_{x^2-y^2}|L_z|d_{xy} \rangle|^2G_{\rm Co}^{\downarrow\downarrow}$ and $|\langle d_{yz}|L_z|d_{zx} \rangle|^2G_{\rm Co}^{\downarrow\downarrow}$, enhancing the perpendicular components through the Bruno term.
Furthermore, we found large spin-flip matrix elements of the interface Pd, such as $|\langle d_{yz}|L_x|d_{x^2-y^2} \rangle|^2G_{\rm Pd}^{\uparrow\downarrow}$, contributing to the cigar-type distribution of spin density and the perpendicular MCA through   van der Laan    term.

\section{Fe/MgO(001) interface}
Finally, we discuss the MCA of Fe/MgO(001) interfaces in magnetic tunnel junctions (MTJs), which are important for spintronics applications.
Fe/MgO-based MTJs exhibit large tunneling magnetoresistive (TMR) ratios of over 400 \% at room temperature because of coherent tunneling of the highly spin-polarized $\Delta _1$ evanescent states through the MgO barrier \cite{2001Butler-PRB,2004Parkin-NM,2004Yuasa-NM}.
To enhance the thermal stability of magnetization directions at a finite temperature, MTJs with perpendicular magnetization are required. However, because body-centered cubic (bcc) Fe and FeCo alloys have cubic structures in bulk, we cannot expect a large perpendicular MCA in the bulk electrode regions of MTJs. Thus, the perpendicular MCA of the interface plays an important role in stabilizing the magnetization directions while preserving the large TMR ratios.

Maruyama {\it et al}. reported perpendicular magnetization of a Fe/MgO(001) interface and a large voltage-controlled MCA (VCMA) change in a few atomic layers of Fe \cite{2009Maruyama-NN}.
Ikeda {\it et al}. showed both a high TMR ratio of over 120 \% at room temperature and the perpendicular magnetization of CoFeB/MgO/CoFeB(001) MTJs when the CoFeB layer was approximately 1.3 nm thick \cite{2010Ikeda-NM}. 
Furthremore, a large perpendicular MCA of 1.4 MJ/m$^3$ in Fe/MgO(001) was observed for 0.7 nm thick Fe layer with adsorbate-induced surface reconstruction \cite{2013Koo-APL}.
XMCD measurements were also performed for ultrathin Fe/MgO(001), and the orbital moment anisotropy was dominant at the Fe/MgO interface perpendicular to MCA; the contribution of quadrupole moments was small but finite at the lattice distorted interfaces \cite{2019Okabayashi-APL}.

\begin{figure}[tp]
\begin{center}
\includegraphics[height=0.26\textheight,width=0.9\textwidth]{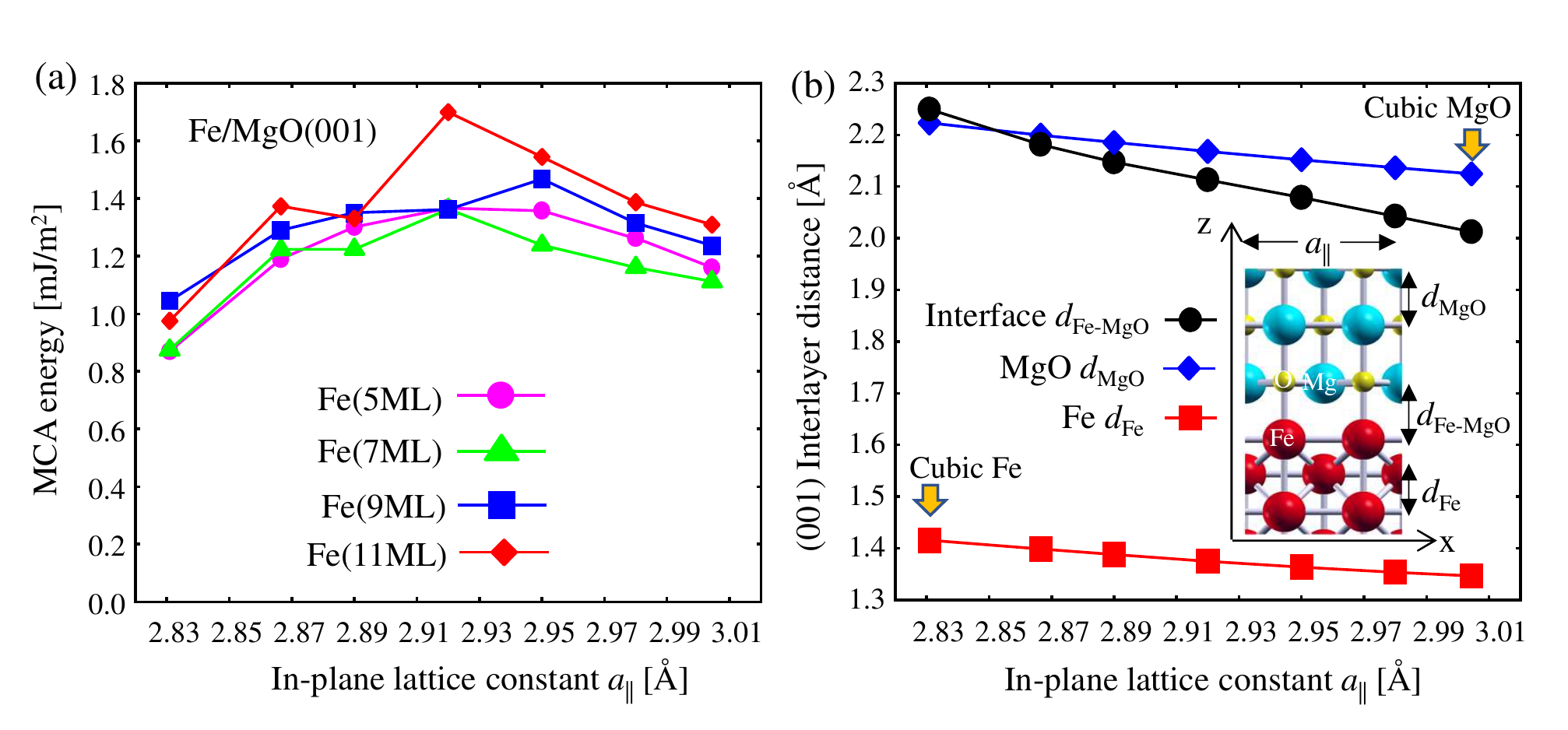}
\end{center}
  \caption{(Color online)(a)The tetragonal distortion dependence of MCA energies of Fe($n_{\rm Fe}$)/MgO(5 ML)(001) for $n_{\rm Fe}$=5, 7, 9 and 11 ML. (b)The optimized (001) interlayer distance of Fe(11 ML)/MgO(5 ML)(001) interface, bulk MgO, and bulk Fe as a function of the in-plane lattice constant $a_{\parallel}$. The atomic structure of Fe/MgO(001) interface is shown in the inset.}
\end{figure}

Several studies have discussed the origin of the perpendicular MCA of Fe/MgO(001) and the VCMA effect in terms of the orbital moment anisotropy and hybridization of the Fe 3$d_{3z^2-r^2}$ orbital with the O 2$p_z$ orbital \cite{2009Tsujikawa-PRL,2010Niranjan-APL,2010Nakamura-PRB,2011Yang-PRB,2014Yoshikawa-APEX} from a theoretical perspective. 
However, the effects of the quadrupole moments (anisotropy of the spin-density distribution) on the perpendicular MCA and the correlation between the lattice distortion and MCA have not been thoroughly discussed for Fe/MgO(001) \cite{2016Odkhuu-SciRep,2018Masuda-PRB}. 
We calculated the MCA energies of the Fe($n_{\rm Fe}$)/MgO(5 ML)(001) interface  for various layer Fe layer thickness $n_{\rm Fe}$ as a function of the in-plane lattice constant $a_{\parallel}$, where $a_{\parallel}$ was changed from $a_{\rm Fe}=2.8309$ \AA~ to $a_{\rm MgO}/\sqrt{2}=3.0043$ \AA~ with an interval of approximately 0.03 \AA. The values of $a_{\rm Fe}$ and $a_{\rm MgO}$ were obtained using DFT calculations for bulk bcc Fe and rock-salt MgO. In addition, we changed the number of Fe-layer $n_{\rm Fe}$ from 5 ML to 11 ML in Fe/MgO(001) supercells. The out-of-plane atomic distances of the supercells with each $a_{\parallel}$ were fully optimized by DFT calculations. Other details of DFT calculations were the same as those in Ref. \cite{2018Masuda-PRB}.

Figure 9(a) shows  the MCA energy of Fe/MgO(001) interface as a function of $a_{\parallel}$.
First, we found that the MCA energies of Fe/MgO(001) exhibited similar $a_{\parallel}$ dependences, irrespective of $n_{\rm Fe}$, although there were some differences. We confirmed that the Fe/MgO(001) supercells with $n_{\rm Fe}\leqq4$ ML showed totally different $a_{\parallel}$ dependence compared to that with $n_{\rm Fe}\geqq5$ ML in Fig. 9(a). This indicates that $n_{\rm Fe}\geqq5$ ML is necessary to correctly describe the characteristics of the Fe/MgO(001) interface. The MCA energy of Fe/MgO(001) in Fig. 9(a) exhibits a non-monotonic behavior with respect to the change of the in-plane lattice constant; it gradually increases and reaches a maximum around $a_{\parallel}=2.92$ \AA~ and then gradually decreases with increasing $a_{\parallel}$. This result is consistent with the previous calculation results \cite{2016Odkhuu-SciRep}.  
Fig. 9(b)  shows the out-of-plane interlayer distance as a function of $a_{\parallel}$ for $n_{\rm Fe}=11$ ML. The result shows a monotonic decrease with increasing $a_{\parallel}$, indicating that the non-monotonic behavior of MCA energies with tetragonal distortion do not originate from the structural properties of Fe/MgO(001). 

\begin{figure}[tp]
\begin{center}
\includegraphics[height=0.26\textheight,width=0.9\textwidth]{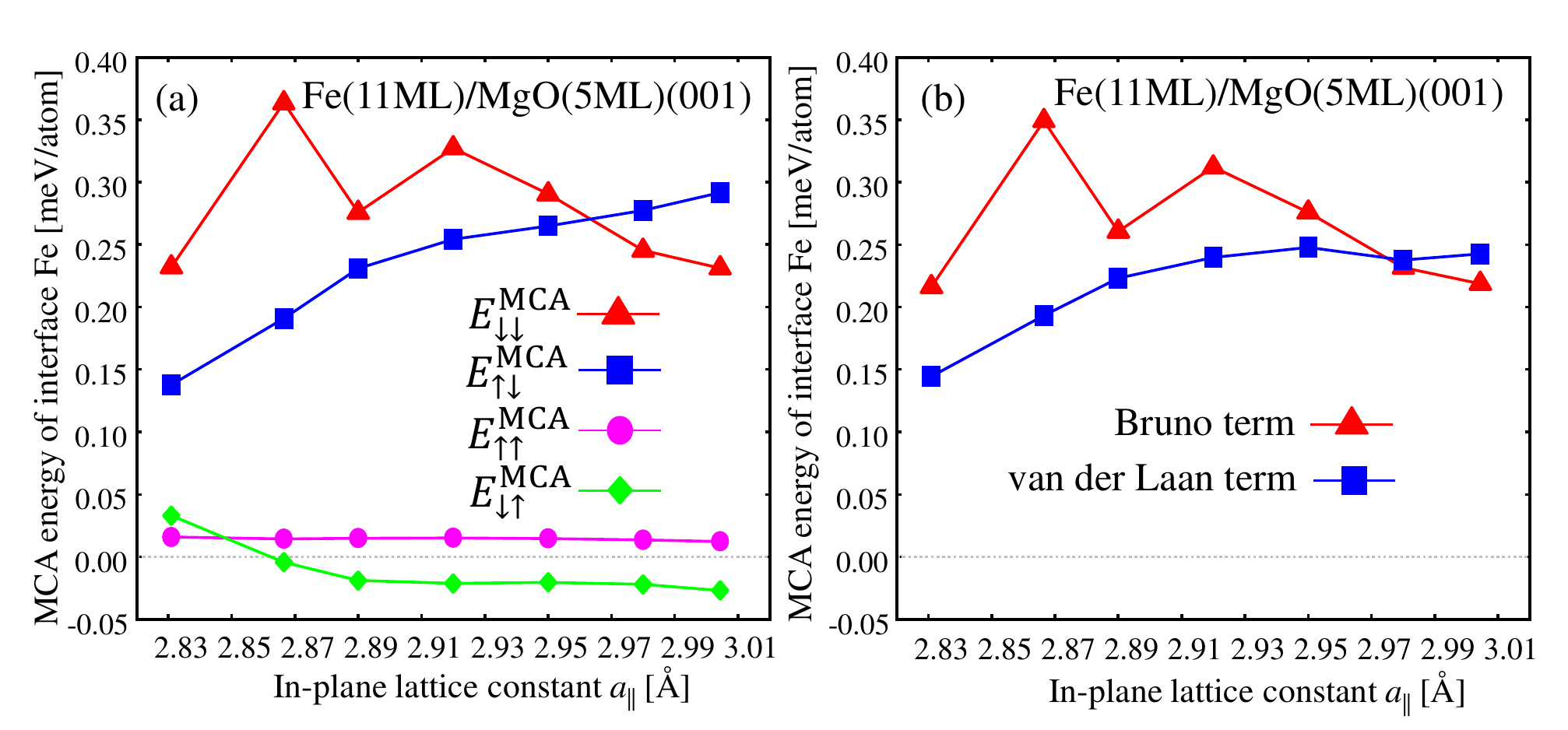}
\end{center}
  \caption{(Color online)(a) Spin-resolved MCA energies in the second-order perturbation of SOI for Fe(11 ML)/MgO(5 ML)(001) interface shown in Eq. (\ref{MCA2}) as a function of $a_{\parallel}$. (b) Bruno term in Eq. (\ref{morbdiff}) and van der Laan term in Eq. (\ref{MCA9}) of Fe/MgO(001) interface as a function of $a_{\parallel}$.}
\end{figure}

To clarify the tetragonal distortion effect on MCA in Fe/MgO(001), we calculated the spin-resolved MCA energies in second-order perturbation of the SOI $E_{\sigma \sigma^{\prime}}^{\rm MCA}$ for the interface Fe atom as a function of the in-plane lattice constant $a_{\parallel}$.
Fig. 10(a) shows that the spin-conserving term $E_{\downarrow \downarrow}^{\rm MCA}$ and spin-flip term $E_{\uparrow \downarrow}^{\rm MCA}$ mainly contribute to the distortion dependence of MCA.
The spin-conserving term $E_{\downarrow \downarrow}^{\rm MCA}$ exhibits non-monotonic behavior for $a_{\parallel}$, whereas the spin-flip term  $E_{\uparrow \downarrow}^{\rm MCA}$ monotonically increases with an increase in $a_{\parallel}$. The spin-conserving term $E_{\downarrow \downarrow}^{\rm MCA}$ is slightly larger than the spin-flip term $E_{\uparrow \downarrow}^{\rm MCA}$ in Fe/MgO(001) interface.

To confirm the relationship  between the MCA energy and the orbital and quadrupole moments, we show in Fig. 10(b) Bruno and   van der Laan    terms of Fe interface as a function of $a_{\parallel}$. To obtain   van der Laan     term, we use $\Delta _{\rm exc}=4$ eV.
Fig. 10(b) shows that the tetragonal distortion dependence of Bruno and   van der Laan    terms are consistent with those of the spin-conserving term $E_{\downarrow \downarrow}^{\rm MCA}$ and the spin-flip term $E_{\uparrow \downarrow}^{\rm MCA}$, respectively. This means that the orbital moment anisotropy and magnetic dipole moment effectively describe the MCA of Fe/MgO(001) with tetragonal distortion.

The increase in   van der Laan     term (the spin-flip term) with increasing $a_{\parallel}$ indicates that Fe/MgO(001) interface has a cigar-type distribution of spin density with tensile distortion, and the perpendicular MCA can be stabilized by orienting the spin moments along the longitudinal direction of the spin-density distributions.
Because the Bruno term (the orbital moment anisotropy) of Fe/MgO(001) for $a_{\parallel}=a_{\rm Fe}$ has approximately the same value as that for $a_{\parallel}=a_{\rm MgO}$, the increase in the perpendicular MCA owing to the tensile tetragonal distortion of Fe/MgO(001) is mainly caused by the additional contribution of the quadrupole moment of spin density around the Fe interface. 
The XMCD and XMLD measurements for Fe/MgO(001) reported that the orbital moment anisotropy was dominant in Fe/MgO(001), with a finite contribution of the quadrupole moment of spin density.

Our calculation results  suggest that the contribution of the anisotropy of quadrupole moment is enhanced when the in-plane lattice constant is close to that of MgO.
The contribution of the spin-flip term in $E_{\rm MCA}$ is expected to be sensitive not only to the in-plane lattice constant, but also to the applied electric field. As a next step, we intend to clarify the relationship between the MCA and lattice distortion, together with the VCMA effects at various magnetic interfaces.

\section{Discussions}

Considering above case studies, we discuss the modulation of orbital moments by local strain. The orbital moments are strongly affected to the anisotropic local environments in the nearest sites through the hybridization. In the case of highly symmetric cubic structure, orbital moments are completely quenched. When the stress can be applied along some direction, the orbital hybridization along this strain direction is preferentially modulated. Within the orbital sum rule in XMCD, the electron occupation in 3d states which is modulated by SOI can be controlled by external strain, resulting in the modulation of orbital moments \cite{2019Okabayashi-Springer}. Therefore, the relationship between strain and orbital moments can be systematically formulated. Our studies in this review generalize the orbital control by strain for some cases. 

   Until now, magneto-striction or magneto-elastic effect has been recognized as strain effect in magnetism as a phenomenologically macroscopic understanding for some materials which are quite essential for applications such as motor, actuator, and mobile devices \cite{2005Chikazumi-Oxford}. However, the microscopic understanding considering the electronic structures and orbital states has not been clarified yet. Recent studies using ultra-thin films strongly require more detailed analysis for strictive effect. Now, we develop novel concept of orbital-strictive or orbital-elastic effect using the results of previous Sections. 

    In the case of Ni/Cu multilayers in Sec. 3, the orbital moments are formulated as a linear relation with strain, which can be understood as orbital-elastic effect. A linear relation originated from the enhancement of orbital moments in the spin-conserving electron motion within the in-plane direction at the interfaces, which is categorized as the case of Fig. 1(a). As shown in Fig. 2(a), orbital moments are related to in-plane strain, resulting in the perpendicular MCA energy. In this case, the direction of strain and enhanced orbital moments are orthogonal in principle. Similar scenario can be adopted to the perpendicular MCA in Co-ferrite CoFe$_2$O$_4$ as orbital-elastic effect in Co$^{2+}$ site with large orbital moment \cite{2022Okabayashi-PRB}. Second type can be understood as quadrupole cases; the direction of strain becomes an easy axis. In this case, the contribution from orbital moments is small and the charge distribution along elongated direction stabilizes the MA. The relation between strain and MCA does not exhibit a linear and spin-flip contribution is dominant for the MCA, which is typical in the case of Mn$_{3-\delta}$Ga because of the small contribution of orbital moment in Mn compounds. These two-types of orbital-elastic effects can be categorized as a microscopic origin of strictive phenomena from the viewpoints of the electronic structures. 

    Other cases for MCA can be also proposed using asymmetric Rashba-type SOI \cite{2014Barnes-SciRep}. For example, Au(111) surface has a strong Rashba-type SOI, which induces the MCA on the deposited ultra-thin Fe layer. Since the potential profile between Fe/Au is different, the interfacial electric gradient promotes the change of charge distributions, which is also detected by XMCD and other spectroscopic probes \cite{2021Okabayshi-PRB,2021Okabayashi-HypInt}. Therefore, the MCA from asymmetric momentum space can be also developed as other orbital-strictive phenomena including a topological physics.

  As discussed in Ref. \cite{1995Weller-PRL} and \cite{2022Cinal-PRB}, the of MCA energy obtained by the first-principles calculations often requires some scaling factors to compare the experimental results. This can be attributed to many factors related to problems both theoretical and experimental points of view. In the calculations, the Bruno term requires to neglect the majority-spin spin-conserving term and the spin-flip terms, which leads to the requirement of the scaling factors. In addition, the first-principles calculations do not include the second Hund's law, which leads to the underestimation of the orbital moment \cite{2013Miura-JPCM}. Furthermore, in the relationship between the spin-flip MCA term and the magnetic dipole moment, the exchange splitting $\Delta _{\rm exc}$ acts as the scaling factor. On the other hand, in the experimental points of view, the MCA energy tends to be smaller than the theoretical predictions due to problems of crystallinity, interface roughness and degree of order of magnetic thin films, leading to the scaling factors in the Bruno and van der Laan terms.

\section{Summary }
In this review, we discussed the perpendicular MCA of magnetic materials and their interfaces based on the orbital and quadrupole moments.
First, we reviewed the relationship among the orbital moments, quadrupole moments, and MCA based on the detailed formulation of the second-order perturbation of the spin-orbit interaction.
We argued that the orbital moment stabilizes the spin moment parallel to the direction with a larger orbital moment, whereas the quadrupole moment stabilizes the spin moment along the longitudinal direction of the spin-density distribution. These effects are expressed as the spin-conserving Bruno term (related to the orbital moment anisotropy) and the spin-flip   van der Laan    term (related to the anisotropy of the quadrupole moment and the magnetic dipole moment). 
We demonstrated that the contributions of $d$ orbitals to these two effects are mutually opposite. In the Bruno term, in-plane $d$ orbitals, such as $d_{x^2-y^2}$ and $d_{xy}$, provide a perpendicular MCA, whereas out-of-plane $d$ orbitals, such as $d_{3z^2-r^2}$, prefer an in-plane MCA. In contrast, in   van der Laan     term, the out-of-plane and in-plane $d$ orbitals provide a perpendicular MCA and an in-plane MCA, respectively. 
The MCA of magnetic materials and interfaces with TMs can be determined from the competition between these two contributions.  

We then applied our formulations of MCA to various magnetic systems by comparing the theoretical results with the XMCD and XMLD measurements.
We showed that fcc Ni with tensile tetragonal distortion shows perpendicular MCA arising from the Bruno term (the orbital moment anisotropy), while MnGa alloys such as L1$_0$-MnGa and D0$_{22}$-Mn$_3$Ga can be attributed to van der Laan term (the  quadrupole moment). Furthermore, the MCA of magnetic systems with interfaces was discussed. We found that the perpendicular MCA of the Co/Pd(111) multilayer originates from the orbital moment anisotropy of the interface Co and the anisotropy of the quadrupole moment at the interface Pd.
Finally, we examined the tetragonal distortion dependence of MCA for Fe/MgO(001) interfaces, which are important systems as the perpendicularly magnetized MTJs with high TMR ratios.
The perpendicular MCA of Fe/MgO(001) exhibited a non-monotonic behavior with respect to the in-plane lattice constant, which can be attributed to both Bruno term (the orbital moment anisotropy) and   van der Laan    term (the quadrupole moment of spin density).
These fundamental understandings of MCA will be essential in the theoretical design of novel magnetic materials and interfaces, as well as for the control of MCA through lattice distortion and applied bias voltage .

\section*{Acknowledgements}
We are grateful to S. Mitani, H. Sukegawa, and K. Masuda at NIMS, H. Yanagihara at the University of Tsukuba, Y. Kota at Fukushima College, M. Shirai and A. Sakuma at Tohoku University for their valuable discussions on our work. For Sec. 3-5, we acknowledge several collaborators, H. Munekata at Tokyo Institute of Technology, T. Taniyama at Nagoya University, S. Mizukami and K. Z. Suzuki at Tohoku University  who prepared excellent samples and provided faithful comments. YM sincerely thanks Y. Suzuki of Osaka University for showing calculation notes regarding detailed derivations of the equations in Ref. [55]. This work was partly supported by the Grants-in-Aid for Scientific Research (Grant Nos. JP16H06332, JP20H00299, JP20H02190, and JP22H04966) from the Japan Society for the Promotion of Science(JSPS), Center for Spintronics Research Network (CSRN) of Osaka University, and Cooperative Research Project Program of the Research Institute of Electrical Communication (RIEC), Tohoku University.


\begin{thebibliography}{10}
\bibitem{explain}
{In the derivation of Bruno term and van der Laan term, we assume that the magnetic materials with more-than-half elements such as Fe, Co, and Ni have fully occupied majority-spin states. In the calculation of the orbital and the magnetic dipole moment, however, we consider the unoccupied majority-spin states. Thus, the effects of partial occupation of the majority-spin state of Fe, Co and Ni are included in the orbital and quadrupole moments of spin density.}
\end{thebibliography}
\end{document}